\documentclass[aip,jcp,reprint,groupedaddress]{revtex4-1}

\pdfoutput=1

 \usepackage{amsmath}
 \usepackage{amsfonts}
 \usepackage{graphicx}
 \usepackage{newtxtext,newtxmath}
 \usepackage{comment}
\usepackage{dcolumn}
\usepackage{natbib}

 \usepackage{hyperref}
 \hypersetup{
    colorlinks=true,       % false: boxed links; true: colored links
    linkcolor=blue,          % color of internal links
    citecolor=blue,        % color of links to bibliography
    filecolor=magenta,      % color of file links
    urlcolor=blue           % color of external links
}

\newcolumntype{d}[1]{D{.}{.}{#1}}

\let\originalleft\left
\let\originalright\right
\renewcommand{\left}{\mathopen{}\mathclose\bgroup\originalleft}
\renewcommand{\right}{\aftergroup\egroup\originalright}

\newcommand{\Zeff}{Z_\text{eff}}
\renewcommand{\to}{\rightarrow}

\begin{document}

\frenchspacing

\title{Effect of molecular constitution and conformation on positron binding and annihilation in alkanes}
\author{A.~R. Swann}
\email{a.swann@qub.ac.uk}
\author{G.~F. Gribakin}
\email{g.gribakin@qub.ac.uk}
\affiliation{
School of Mathematics and Physics, Queen's University Belfast, University Road, Belfast BT7 1NN, United Kingdom}
\date{\today}

\begin{abstract}
The model-potential approach previously developed by the authors to study positron interactions with molecules is used to calculate the positron binding energy for $n$-alkanes (C$_n$H$_{2n+2}$) and the corresponding cycloalkanes (C$_n$H$_{2n}$). For $n$-alkanes, the dependence of the binding energy on the conformation of the molecule is investigated, with more compact structures showing greater binding energies. As a result, thermally averaged binding energies for larger alkanes ($n\gtrsim 9$) show a strong temperature dependence in the range of 100--600~K. This suggests that positron resonant annihilation can be used as a probe of rotational (\textit{trans-gauche}) isomerization of $n$-alkanes. In particular, the presence of different conformers leads to shifts and broadening of vibrational Feshbach resonances in the annihilation rate, as observed with a trap-based low-energy positron beam.
\end{abstract}

\maketitle

\section{\label{sec:intro}Introduction}

Since the initial prediction\cite{Dirac31} and subsequent discovery\cite{Anderson33} of the positron ($e^+$) almost 90 years ago, it has found practical uses in many areas of science, e.g., in fundamental tests of QED and the Standard Model,\cite{Karshenboim05,Ishida14,ALEPH06} astrophysics,\cite{Guessoum14} condensed-matter physics,\cite{Tuomisto13} and medicine.\cite{Wahl02} In this paper, we show that positrons are also sensitive probes of molecular structure. In particular, their binding energies for larger alkanes show strong dependence on molecular conformation.

Positron annihilation rates in polyatomic molecular gases are strongly enhanced compared with the basic Dirac annihilation rate, due to positron attachment into vibrational Feshbach resonances (VFRs).\cite{Gribakin10} This process is possible for molecules that support a bound state for a positron: the incident positron is captured into the bound state, with the excess energy being transferred into molecular vibrations, typically those of a mode with near-resonant energy.\cite{Gribakin00,Gribakin01} 
This leads to pronounced peaks in the positron-energy-resolved annihilation rate.\cite{Gilbert02} The difference between the energy  of the vibrational mode  and the energy  of the peak is a measure of the positron binding energy $\varepsilon_b$, viz.,
\begin{equation}
\varepsilon_b = \hbar\omega_\nu - \varepsilon_\nu , \label{eq:VFR}
\end{equation}
where $\omega_\nu$ is the frequency of vibrational mode $\nu$ and $\varepsilon_\nu$ is the energy of the resonant peak. Positron binding energies have now been measured for over 80 molecules by the Surko group in San Diego.\cite{Barnes03,Barnes06,Young07,Young08,Young08a,Danielson09,Danielson10,Danielson12,Natisin:thesis}

On the side of theory, accurate calculations of positron binding to molecules have proven to be difficult. Most \textit{ab initio} calculations have been for strongly polar molecules, where the existence of a bound state is guaranteed even at the static, Hartree-Fock level of theory.\footnote{A molecule with a dipole moment greater than 1.625~D always has a bound state for a positron (or an additional electron),\cite{Crawford67} although this critical dipole moment increases if the molecule is rotating.\cite{Garrett71}}
Notably, the binding energy increases significantly when electron-positron correlations are included, e.g., for acetonitrile CH$_3$CN, $\varepsilon_b$ increases from 15~meV (Hartree-Fock) to 135~meV (configuration interaction).\cite{Tachikawa11} However, experimental measurements have mostly been for weakly polar and nonpolar molecules, for which \textit{ab initio} calculations have failed to predict binding reliably.\cite{Sugiura19}

We recently proposed a simple physical model to enable calculations of positron binding for a wide range of molecules with predictive capability.\cite{Swann18,Swann19,Swann20}
The molecular geometry is optimized at the Hartree-Fock level using the 6--311++G$(d,p)$ basis, and the electronic molecular orbitals are found and used to obtain the electrostatic potential $V_\text{st}$ of the molecule. Then a potential $V_\text{cor}$ that describes long-range polarization of the molecule by the positron, viz.,
\begin{equation}\label{eq:Vcor}
V_\text{cor}(\mathbf r) = - \sum_A \frac{\alpha_A}{2 \lvert \mathbf r - \mathbf r_A \rvert^4} [1 - \exp(-\lvert\mathbf r - \mathbf r_A \rvert^6/\rho_A^6)] ,
\end{equation}
is added. Here, the sum is over the molecule's constituent atoms $A$, $\mathbf r$ is the position of the positron, and $\mathbf r_A$ is the position of nucleus $A$, relative to an arbitrary origin. Atomic units (a.u.) have been used. The atomic hybrid polarizabilities $\alpha_A$ take into account the chemical environment of each atom within the molecule,\cite{Miller90} and the total polarizability of the molecule is $\alpha=\sum_A\alpha_A$. In Eq.~(\ref{eq:Vcor}), the factor in brackets provides a cutoff of the polarization potential at distances close to an atomic nucleus, parametrized by the cutoff radius $\rho_A$. Its values are chosen to fit an experimentally measured binding energy for a representative molecule, or by comparison with high-quality calculations, if available. The short-range part of $V_\text{cor}$ accounts for other attractive correlation effects, such as virtual positronium formation. The Schr\"odinger equation for the total potential $V_\text{st}+V_\text{cor}$ is solved to obtain the positron binding energy $\varepsilon_b$; in practice, this is done using \textsc{gamess}\cite{Schmidt93,Gordon05} with the \textsc{neo} plugin, \cite{Webb02,Adamson08} which we have modified to include $V_\text{cor}$.\cite{Swann18}

We have previously  used this method to calculate the binding energy and electron-positron contact density for hydrogen cyanide HCN\cite{Swann18} and for  alkane molecules with up to $16$ carbon atoms.\cite{Swann19} We also  investigated elastic scattering of positrons by several atoms and diatomic molecules and calculated the normalized annihilation rate $Z_\text{eff}$ for low-energy positrons.\cite{Swann20}

In this work we return to the problem of positron binding to alkane molecules.
In Ref.~\onlinecite{Swann19}, our calculations for cyclopropane C$_3$H$_6$ and cyclohexane C$_6$H$_{12}$ found  their positron binding energies to be smaller than that of $n$-propane C$_3$H$_8$ and $n$-hexane C$_6$H$_{14}$, respectively. This was explained as a result of each cycloalkane having two fewer hydrogen atoms than the corresponding $n$-alkane, and, hence, a smaller dipole polarizability and a smaller value of $\varepsilon_b$. We also found that the positron binding energies for the three structural isomers of pentane, viz., $n$-pentane, isopentane, and neopentane, were close. This was explained as a result of the three structural isomers having the same constitution, and, hence, the same polarizability. It thus appeared that for smaller alkanes, the dipole polarizability was the main parameter that determined the strength of binding (in agreement with the empirical scaling found in Ref.~\onlinecite{Danielson09}), with short-range structural effects playing a relatively small role.

For the $n$-alkane sequence, we found that the growth of $\varepsilon_b$ with $n$ slows for larger values of $n$, i.e., the binding energy begins to ``level off'' for sufficiently large values of $n$.\cite{Swann19} In contrast, the experimental data indicate that the binding energy continues to increase with  $n$ in a near-linear fashion, at least up to $n=16$.\cite{Young08} Our calculations assumed that the molecules were in the lowest-energy extended (all-\textit{trans}) conformation, and we tentatively suggested that the discrepancy with the experimental data could be due to such large chain molecules favoring other conformations  at room temperature.\cite{Swann19,Thomas06}

Here we investigate the dependence of the positron binding energy on the molecular constitution and conformation for several alkane molecules.
Firstly, we consider the positron binding energy as a function of the molecular polarizability for cycloalkanes C$_n$H$_{2n}$ and $n$-alkanes C$_n$H$_{2n+2}$ up to $n=10$. We show that for $n\leq 6$, the value of $\varepsilon_b$ is determined almost entirely by the dipole polarizability, while for $n\geq7$, the constitution of the molecule also plays a significant role. Secondly, we calculate positron binding energies for the possible conformers of several $n$-alkanes up to $n=16$ and provide expectation values of $\varepsilon_b$ for ensembles of room-temperature molecules. 
We also investigate the temperature dependence of the expected positron binding energy and the effect of the presence of multiple conformers in the gas  on the measured annihilation rate.
As in Ref.~\onlinecite{Swann19}, we take $\alpha_\text{C}=7.096$~a.u., $\alpha_\text{H}=2.650$~a.u.,  $\rho_\text{C}=\rho_\text{H}=2.25$~a.u., and we solve the Schr\"odinger equation for the positron using an even-tempered Gaussian basis consisting of 12 $s$-type primitives centered on each C nucleus, with exponents $0.0001\times3^{i-1}$ ($i=1$--12), and eight $s$-type primitives centered on each H nucleus, with exponents $0.0081\times3^{i-1}$ ($i=1$--8).

%The paper is structured as follows. In Sec.~\ref{enumeration} we outline the labeling and enumeration of the possible conformers of $n$-alkane molecules. In Sec.~\ref{positron_theory} we describe the salient features of our model for calculating positron-molecule binding energies, previously used in Refs.~\onlinecite{Swann18,Swann19,Swann20}. We present the main results in Sec.~\ref{results}, namely, the calculated average positron binding energies for several $n$-alkane molecules when the presence of conformers is taken into account. We compare these results with our earlier calculations for the straight-chain molecules\cite{Swann19} and with experimental data.\cite{Young08}

\section{\label{sec:structure}Effect of molecular constitution  on the positron binding energy}

\begin{figure}
\centering
\includegraphics[width=\columnwidth]{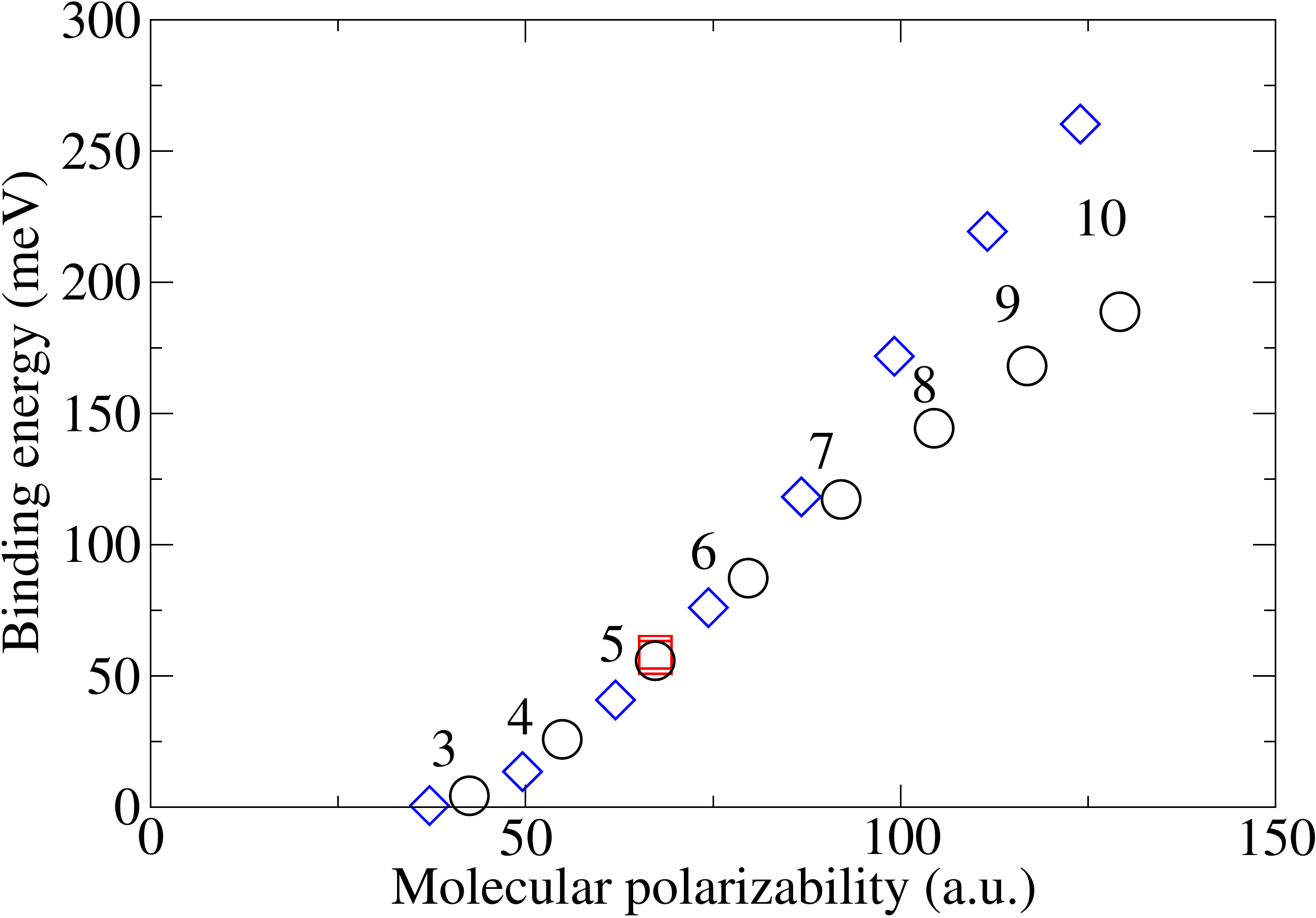}
\caption{\label{fig:cyclo-vs-n}Positron binding energy as a function of the molecular polarizability for alkanes up to $n=10$. Black circles, straight-chain $n$-alkanes;\cite{Swann19} blue diamonds, cycloalkanes; red squares, isopentane and neopentane.\cite{Swann19} Labels indicate the value of $n$ for each $n$-alkane-and-cycloalkane pair.}
\end{figure}

To supplement the values of $\varepsilon_b$ for cyclopropane C$_3$H$_6$ and cyclohexane C$_6$H$_{12}$ calculated in Ref.~\onlinecite{Swann19}, we have calculated $\varepsilon_b$ for the other cycloalkanes C$_n$H$_{2n}$ up to $n=10$. 
Figure~\ref{fig:cyclo-vs-n} shows these binding energies, along with those of the straight-chain (all-\textit{trans}) $n$-alkanes,\cite{Swann19} as a function of the molecular polarizability. For completeness, we also show the previously calculated binding energies of isopentane and neopentane.\cite{Swann19} The corresponding numerical values are listed in Table~\ref{tab:constitution_binen}.

\begin{table}
\caption{\label{tab:constitution_binen}Positron binding energies for $n$-alkanes C$_n$H$_{2n+2}$\cite{Swann19} and cycloalkanes C$_n$H$_{2n}$, with their molecular polarizabilities $\alpha$, for $n=3$--10. The binding energies for isopentane and neopentane are also shown.\cite{Swann19}}
\begin{ruledtabular}
\begin{tabular}{lllll}
 & \multicolumn{2}{c}{C$_n$H$_{2n}$}  & \multicolumn{2}{c}{C$_n$H$_{2n+2}$} \\
 \cline{2-3} \cline{4-5}
$n$ & $\alpha$ (a.u.) & $\varepsilon_b$ (meV) & $\alpha$ (a.u.) & $\varepsilon_b$ (meV) \\
\hline
3 & \phantom{0}37.19 & \phantom{00}0.5521 & \phantom{0}42.49 & \phantom{00}4.302 \\
4 & \phantom{0}49.58 & \phantom{0}13.48 & \phantom{0}54.88 & \phantom{0}25.81 \\
5 & \phantom{0}61.98 & \phantom{0}40.78 & \phantom{0}67.28 & \phantom{0}55.75\footnotemark[1] \\
&&&& \phantom{0}58.91\footnotemark[2] \\
&&&& \phantom{0}57.40\footnotemark[3] \\
6 & \phantom{0}74.38 & \phantom{0}75.62 & \phantom{0}79.68 & \phantom{0}87.23 \\
7 & \phantom{0}86.77 & 118.2 & \phantom{0}92.07 & 117.2 \\
8 & \phantom{0}99.17 & 171.8 & 104.5 & 144.4 \\
9 & 111.6 & 219.4 & 116.9 & 168.1 \\
10 & 124.0 & 260.3 & 129.3 & 188.8
\end{tabular}
\end{ruledtabular}
\footnotetext[1]{$n$-Pentane.}
\footnotetext[2]{Isopentane.}
\footnotetext[3]{Neopentane.}
\end{table}

For $n\leq6$, we see that the positron binding energy is determined by the molecular polarizability alone, and the value of $\varepsilon_b$ for each cycloalkane is lower than that for the corresponding $n$-alkane due to the lower polarizability. However, cycloheptane and $n$-heptane ($n=7$) have almost the same value of $\varepsilon_b$. For $n\geq8$, the value of $\varepsilon_b$ for the cycloalkane is significantly larger than that for the corresponding $n$-alkane, despite the lower polarizability. These observations can be explained as follows.
For small alkanes ($n\leq6$), the positron binding energy is small ($\varepsilon_b<100$~meV), and the characteristic extent of the positron wavefunction $r_p\sim 1/\sqrt{2\varepsilon _b}$ (in atomic units) is greater than the size of the molecule.\footnote{The wavefunction of a bound state in a short-range potential decreases asymptotically as $e^{-\kappa r}/r$, so $r_p\sim 1/\kappa $, where $\kappa = \sqrt{2\varepsilon _b}$ in atomic units.} The positron is thus found mostly at large distances from the molecule, where its wave function remains largely spherical. Hence, the positron binding is sensitive to the asymptotic form of the polarization potential $V_\text{cor}$, and the precise geometry of the molecule at short range does not play a big role. Conversely, for larger alkanes ($n\geq7$), the spatial extent of the positron wavefunction becomes smaller than the length of the $n$-alkane carbon backbone. Its wavefunction becomes elliptical, stretched along the molecule (see Fig. 2 in Ref.~\onlinecite{Swann19}), and the positron no longer interacts equally strongly with all the atoms in the molecule. For the more compact, cyclic form, however, the advantage of ``sampling'' all the atoms is largely preserved, leading to a stronger overall positron-molecule attraction.
This can be seen in Fig.~\ref{fig:C10H20_pso0,017}, which shows the shape of the bound positron wave function for cyclodecane C$_{10}$H$_{20}$ (the wave function has a value of 0.017~a.u. on the surface shown).
\begin{figure}
\centering
\includegraphics[width=0.95\columnwidth]{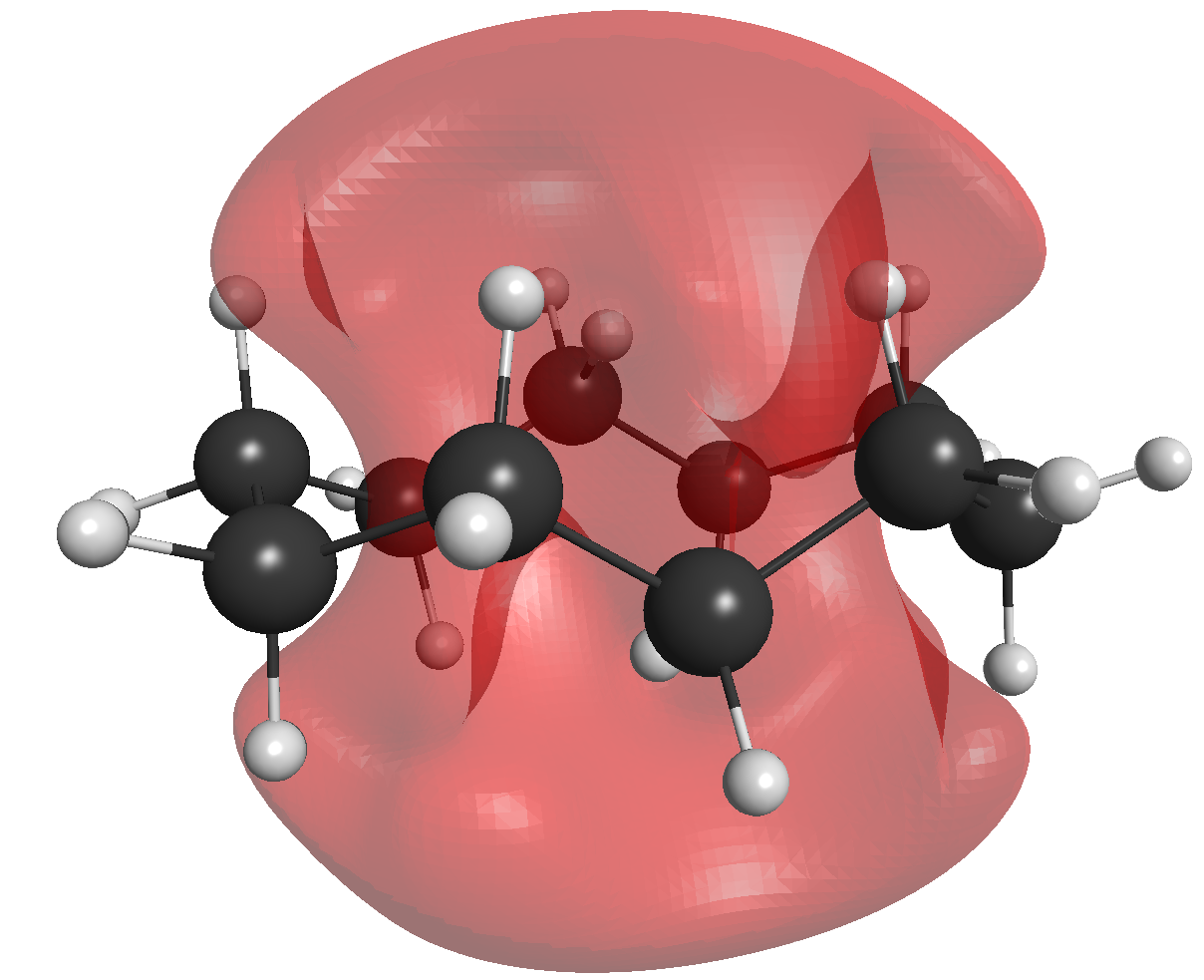}
\caption{\label{fig:C10H20_pso0,017}Wave function of the positron bound state for cyclodecane C$_{10}$H$_{20}$. The wave function has a value of 0.017~a.u. on the surface shown.}
\end{figure}
As a result, the cycloalkanes with $n>7$ have greater positron binding energies than the corresponding $n$-alkanes, despite having fewer atoms and smaller dipole polarizabilities.

This observation indicates that positron binding energies for nonextended conformations of larger alkanes can be noticeably greater than those of the straight-chain structures. Such conformations dominate the isomer distribution for long alkane chains at room temperature. This can lead to temperature-dependent measured values of the positron binding energy and other effects of temperature on the positron-molecule resonant annihilation rates. We address these questions in the following, main section of the paper.

\section{\label{sec:conformation}Effect of molecular conformation on the positron binding energy and annihilation rate}

\subsection{Room-temperature average binding energies}\label{subsec:av_be}

\begin{figure}
\centering
\includegraphics[width=0.95\columnwidth]{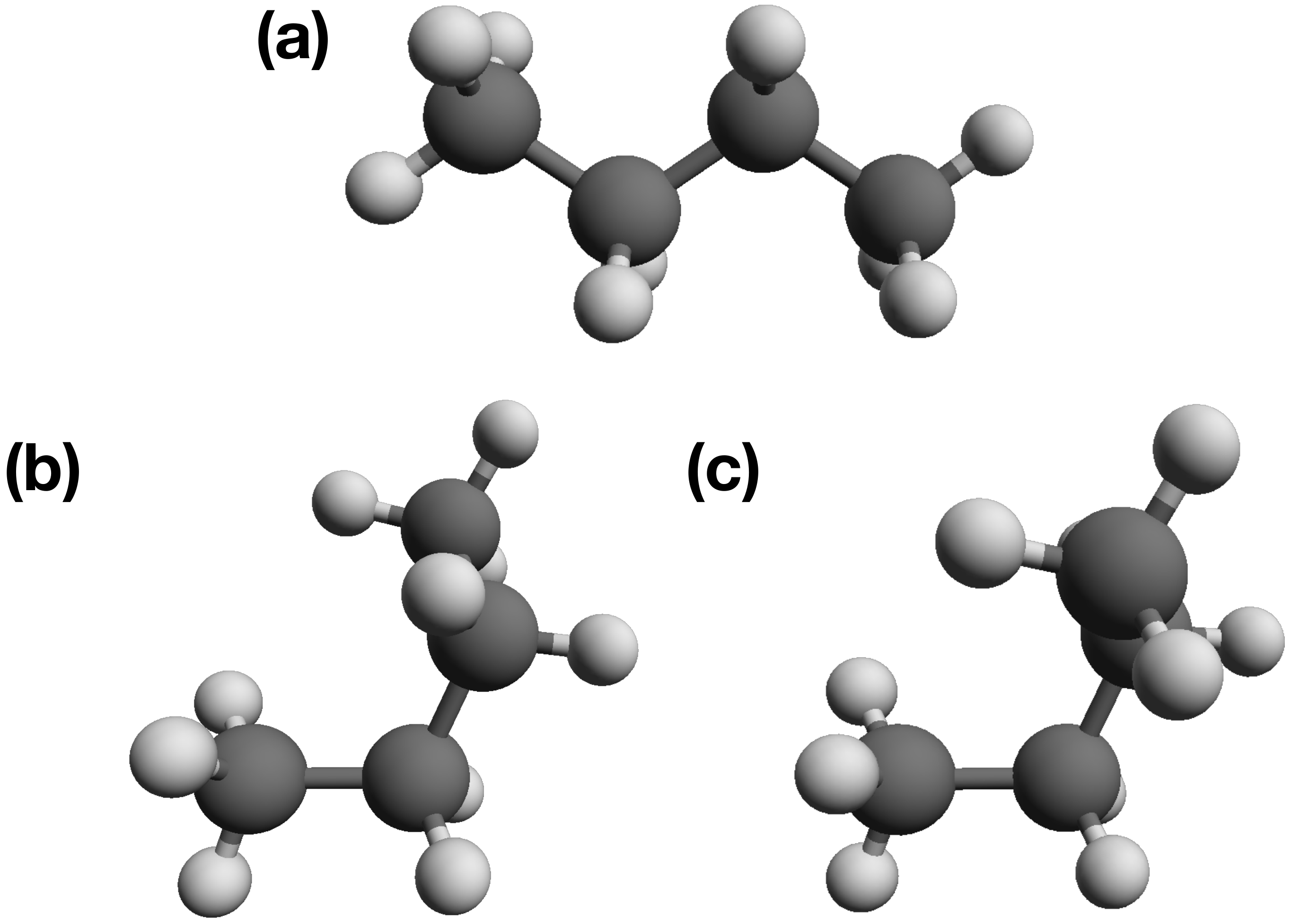} 
\caption{\label{fig:butane-conformers}(a) \textit{trans}, (b) \textit{gauche}$^+$, and (c) \textit{gauche}$^-$ conformers of $n$-butane.}
\end{figure}

Conformers of $n$-alkanes for $n\geq4$ can be classified using the labels \textit{trans} (also known as \textit{anti}), denoted $t$, and \textit{gauche}, denoted $g$. These labels are used to specify the torsional angle of each successive C--C bond following the initial two bonds: a torsional (or dihedral) angle of ${\approx}180^\circ$ is labeled $t$, while a torsional angle of ${\approx}60^\circ$ is labeled $g$. Note that a torsional angle of ${\approx}60^\circ$ can be counterclockwise or clockwise, so we denote these as $g^+$ and $g^-$, respectively.
Figure \ref{fig:butane-conformers} shows the $t$, $g^+$, and $g^-$ conformers of $n$-butane. 
The $g^-$ conformer is an enantiomer of the $g^+$ conformer, and consequently the two are spectroscopically indistinguishable.
As a second example, the possible conformers  of $n$-pentane are $tt$, $tg^+$, $tg^-$, $g^+g^+$,  $g^-g^-$, and $g^+g^-$.\footnote{The $g^+g^-$ conformer is sterically strained and is sometimes denoted $g^+x^-$, where $x$ denotes a ``perpendicular'' structure. The torsional angle of an $x^\pm$ C--C bond is ${\approx}95^\circ$.\cite{Csontos16} This effect is known as pentane interference.} At low temperatures, linear alkanes of moderate length prefer a fully extended, lowest-energy \textit{all-trans} conformation. (For $n>17$--18, however, a hairpin structure may become lower in energy due to self-solvation effects.\cite{Luettschwager13})

At thermal equilibrium, the fraction $p_i$ of a particular conformer $i$ present in a gas sample of an $n$-alkane is determined by a Boltzmann distribution, viz.,
\begin{equation}
p_i = \frac{w_i e^{-E_i / k_BT}}{\sum_j w_j e^{-E_j / k_BT}} , \label{eq:prob-dist}
\end{equation}
where $w_i$ is the degeneracy of the conformer, $E_i$ is the energy of the conformer relative to the lowest-energy all-$t$ conformer, $k_B$ is the Boltzmann constant, $T$ is the temperature, and the sum in the denominator is over all spectroscopically distinguishable conformers.

When positrons annihilate in a gas, their characteristic collision time with a molecule is shorter than the typical lifetimes of various conformations. As a result, they sample the various conformers, and the measured binding energy can be approximated by the average positron binding energy across in the sample,
\begin{equation}
\langle \varepsilon_b \rangle _T= \sum_i p_i \varepsilon_b^{(i)} , \label{eq:av_bine}
\end{equation}
where $\varepsilon_b^{(i)}$ is the positron binding energy for conformer $i$. Equation (\ref{eq:av_bine}) assumes that the probability of positron attachment and annihilation in various conformers is similar. To justify this assumption, we note that the dominant mechanism of positron capture is by excitation of infrared-active modes of the molecule\cite{Gribakin06} (followed by intramolecular vibrational energy redistribution, for larger molecules\cite{Gribakin04}) with infrared-inactive modes producing a distinct but small contribution.\cite{Natisin17}
Of course, the vibrational spectra of the various conformers have some individual features (see, e.g., Ref.~\onlinecite{Luettschwager13}). However, the overall infrared  absorption profiles of $n$-alkanes remain quite similar, with certain infrared strengths that can be assigned to various absorption bands, e.g., asymmetric and symmetric stretches of the C--H bond in CH$_3$ methyl groups and slightly lower frequency C--H stretches of the CH$_2$ groups (around 2800--2900~cm$^{-1}$), deformations of the CH$_3$ and CH$_2$ groups around 1400--1500~cm$^{-1}$, etc.\cite{Sverdlov74} Another possible source of the conformer dependence of the annihilation probability is the difference in the binding energies for various conformers. For larger alkanes this can reach a few tens of meV (see below), but it does not exceed $\sim $10\% of the average, at least at room temperature.

The total number of possible conformers for the general $n$-alkane C$_n$H$_{2n+2}$ is $\sum_iw_i=3^{n-3}$ (since there are $n-3$ dihedral angles that can each be either $t$, $g^+$, or $g^-$). This exponential increase of  the total number of conformers with molecular size makes it impractical to calculate the binding energy $\varepsilon_b^{(i)}$ for all conformers to obtain $\langle \varepsilon_b \rangle _T$ from Eq.~(\ref{eq:av_bine}) for large $n$. Excluding conformers that contain two consecutive oppositely signed \textit{gauche} angles, i.e., $g^+g^-$ or $g^-g^+$, which are energetically unfavorable, reduces the number of binding energies that have to be calculated. In fact, this leaves a total of 
\begin{equation}
\frac12 \left[ \big(1-\sqrt{2}\big)^{n-2} + \big(1 + \sqrt{2}\big)^{n-2} \right] =  \sum_{k=0}^{\lfloor n/2-1 \rfloor }
\frac{(n-2)! \, 2^k}{(2k)! \, (n-2k-2)!} \label{eq:restricted_total}
% \binom{n-2}{2k} 
\end{equation}
conformers (see Appendix \ref{sec:enumeration}). This number also grows exponentially. Table \ref{tab:no_conformers} shows the number of possible conformers for  $n$-alkanes with $n=4$--16, categorized by the number of \textit{gauche} angles $N_g$, excluding conformers with $g^\pm g^\mp$ pairs. Note that conformers that read the same forwards and backwards, e.g., $tg^+t$ and $ttg^-tg^+g^+tg^-tt$, are only counted once, while those that do not, e.g., $ttg^+$ and $g^+ttg^-$, are counted twice.
\begin{table*}
\caption{\label{tab:no_conformers}Number of possible (not necessarily spectroscopically distinct) conformers for each $n$-alkane molecule with a fixed number $N_g$ of \textit{gauche} angles. Conformers with adjacent $g^+$ and $g^-$ angles are not included.}
\begin{ruledtabular}
\begin{tabular}{rrrrrrrrrrrrrrrr}
             & \multicolumn{14}{c}{$N_g$} & \\
             \cline{2-15}
% \mc{$n$}     & \mc{0} & \mc{1} & \mc{2} & \mc{3} & \mc{4} & \mc{5} & \mc{6} & \mc{7} & \mc{8} & \mc{9} & \mc{10} & \mc{11} & \mc{12} & \mc{13} & \mc{total} \\
$n$ & 0 & 1 & 2 & 3 & 4 & 5 & 6 & 7 & 8 & 9 & 10 & 11 & 12 & 13 & total \\
\hline
4 & 1 & 2 &&&&&&&&&&&&& 3    \\
5 & 1 & 4 & 2 &&&&&&&&&&&& 7   \\
6 & 1 & 6 & 8 & 2  &&&&&&&&&&& 17   \\
7 & 1 & 8 & 18 & 12 & 2 &&&&&&&&&& 41     \\
8 & 1 & 10 & 32 & 38 & 16 & 2 &&&&&&&&& 99    \\ 
9 & 1 & 12 & 50 & 88 & 66 & 20 & 2 &&&&&&&& 239       \\
10 & 1 & 14 & 72 & 170 & 192 & 102 & 24 & 2 &&&&&&& 577    \\
11 & 1 & 16 & 98 & 292 & 450 & 360 & 146 & 28 & 2 &&&&&& 1393    \\
12 & 1 & 18 & 128 & 462 & 912 & 1002 & 608 & 198 & 32 & 2 &&&&& 3363     \\
13 & 1 & 20 & 162 & 688 & 1666 & 2364 & 1970 & 952 & 258 & 36 & 2 &&&& 8119    \\
14 & 1 & 22 & 200 & 978 & 2816 & 4942 & 5336 & 3530 & 1408 & 326 & 40 & 2 &&& 19\,601     \\
15 & 1 & 24 & 242 & 1340 & 4482 & 9424 & 12\,642 & 10\,836 & 5890 & 1992 & 402 & 44 & 2 && 47\,321      \\
16  & 1 & 26 & 288 & 1782 & 6800 & 16\,722 & 27\,008 & 28\,814 & 20\,256 & 9290 & 2720 & 486 & 48 & 2 & 114\,243     
\end{tabular}
\end{ruledtabular}
\end{table*}

The positron binding energies need to be calculated only for the conformers that are spectroscopically distinct, i.e., those that do not read the same forwards and backwards, and that are not obtained from each other by mirror symmetry (interchanging $g^-$ and $g^+$). Table \ref{tab:n7conf} lists all spectroscopically distinct conformers for
$n\leq 7$ along with their degeneracies $w_i$.
\begin{table}
\caption{\label{tab:n7conf}Conformers of $n$-alkanes with $n=4$--7, with their degeneracies $w_i$, energies $E_i$ with respect to the all-$t$ conformer, and positron binding energies $\varepsilon_b^{(i)}$.}
\begin{ruledtabular}
\begin{tabular}{cccccc}
$n$ & $N_g^{(i)}$ & Conformer $i$ & $w_i$ & $E_i$ (meV) &  $\varepsilon_b^{(i)}$ (meV) \\
\hline
4 & 0 & $t$ & 1 & 0 & 25.81\footnotemark[1] \\
   & 1 & $g^\pm$ & 2 & 42.37 & 27.63 \\
5 & 0 & $tt$ & 1 & 0 & 55.75\footnotemark[1]  \\
   & 1 & $tg^\pm$ & 4 & 44.21 & 58.56 \\
   & 2 & $g^\pm g^\pm$ & 2 & 82.78 & 62.18 \\
6 & 0 & $ttt$ & 1 & 0 & 87.23\footnotemark[1]  \\
   & 1 & $ttg^\pm$ & 4 & 43.38 & 92.52 \\
   & 1 & $tg^\pm t$ & 2 & 45.60 & 90.77 \\
   & 2 & $tg^\pm g^\pm$ & 4 & 84.29 & 97.45 \\
   & 2 & $g^\pm t g^\pm$ & 2 & 87.13 & 97.17 \\
   & 2 & $g^\pm t g^\mp$ & 2 & 92.08 & 94.95 \\
   & 3 & $g^\pm g^\pm g^\pm$ & 2 & 123.8 & 101.0 \\
7 & 0 & $tttt$ & 1 & 0 & 117.2\footnotemark[1]  \\  
   & 1 & $tttg^\pm$ & 4 & 43.65 & 123.4 \\ 
   & 1 & $ttg^\pm t$ & 4 & 45.18 & 123.8 \\
   & 2 & $ttg^\pm g^\pm$ & 4 & 83.89 & 132.3 \\
   & 2 & $tg^\pm tg^\pm$ & 4 & 88.81 & 129.0 \\
   & 2 & $tg^\pm tg^\mp$ & 4 & 94.32 & 126.4 \\
   & 2 & $g^\pm tt g^\pm$ & 2 & 86.93 & 130.2 \\
   & 2 & $g^\pm tt g^\mp$ & 2 & 88.53 & 132.4 \\
   & 2 & $t g^\pm g^\pm t$ & 2 & 86.15 & 132.7 \\
   & 3 & $tg^\pm g^\pm g^\pm$ & 4 & 125.6 & 137.0 \\
   & 3 & $g^\pm t g^\pm g^\pm$ & 4 & 127.9 & 140.2 \\
   & 3 & $g^\pm t g^\mp g^\mp$ & 4 & 132.9 & 134.6 \\
   & 4 & $g^\pm g^\pm g^\pm g^\pm$ & 2 & 165.1 & 140.2
\end{tabular}
\end{ruledtabular}
\footnotetext[1]{The positron binding energies for the all-$t$ conformers are the same values quoted in Table I of Ref.~\onlinecite{Swann19}.}
\end{table}
It also shows the Hartree-Fock value of $E_i$ for each conformer and the calculated positron binding energy $\varepsilon_b^{(i)}$. The values of $E_i$ are in agreement with the accepted range of the \textit{trans}-\textit{gauche} energy differences, $\Delta E_{tg} = 0.5$--1.0~kcal/mol\cite{Thomas06,Tsuzuki91} (or, equivalently, 22--44~meV). Regarding the binding energies, we observe that they increase with the increase in the number of \textit{gauche} angles. For example, for $n$-pentane, the binding energy grows from 56~meV for $tt$ to 62~meV for $g^\pm g^\pm $. This effect becomes stronger for larger alkanes, e.g.,
the binding energy increases from 87~meV for an all-$t$ to 101~meV for an all-$g$ conformer of $n$-hexane, and from 117~meV for an all-$t$ to 140~meV for an all-$g$ conformer of $n$-heptane. This trend is in agreement with the observation made in Sec.~\ref{sec:structure} that more compact molecular structures lead to greater positron binding.

Using Eqs.~(\ref{eq:prob-dist}) and (\ref{eq:av_bine}), we obtain the following values of the average positron binding energy $\langle \varepsilon_b \rangle _T$ at room temperature ($T=300$~K, $k_BT=26$~meV): $n$-butane, 26.32~meV; $n$-pentane, 57.17~meV; $n$-hexane, 90.62~meV; $n$-heptane, 123.2~meV. These values are 2.0\%, 2.5\%, 3.9\%, and 5.1\% larger than the binding energies of the corresponding all-$t$ conformers. This indicates that more compact higher-energy conformers  (i.e., conformers with a greater number of \textit{gauche} angles) play an increasingly important role in determining the observed positron binding energy in larger alkanes. Here the unfavorable Boltzmann factor $e^{-\Delta E_{tg}/k_BT}\sim 0.2$--0.4 for each extra \textit{gauche} angle is countered by the larger number of such conformers.

\begin{figure}
\centering
\includegraphics[width=\columnwidth]{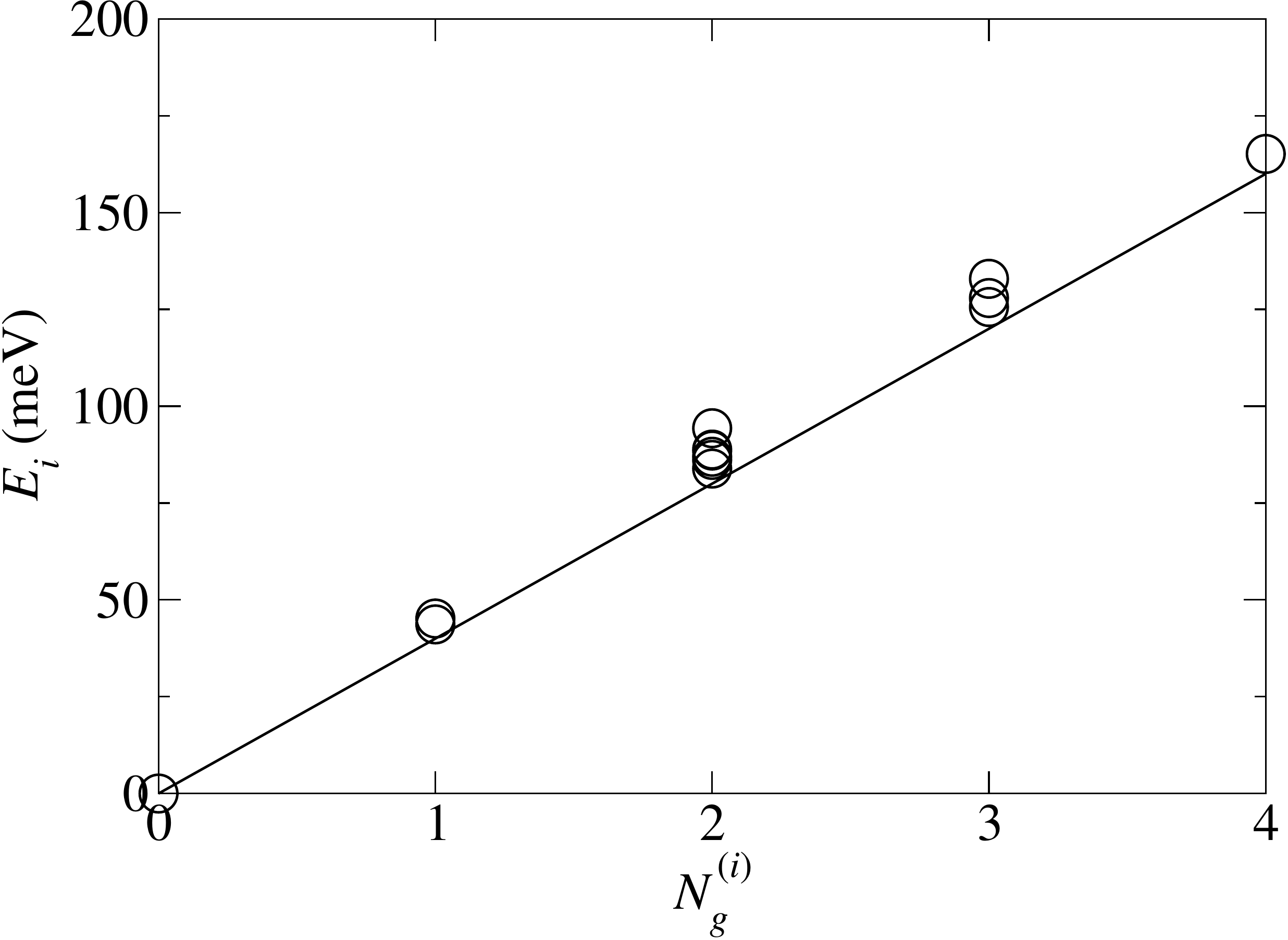}
\caption{\label{fig:heptane_DeltaE}Energies of the conformers of $n$-heptane C$_7$H$_{16}$ calculated at the Hartree-Fock level relative to the all-$t$ conformer, as a function of the number of gauche bonds $N_g^{(i)}$ (circles); line, $E_i = 40 N_g^{(i)}$~meV.}
\end{figure}

For $n\geq8$, there are simply too many spectroscopically different conformers (even without the $g^\pm g^\mp$ pairs) to calculate their positron binding energies individually and to obtain $\langle \varepsilon_b \rangle _T$ from Eq.~(\ref{eq:av_bine}). We therefore explore the role of conformations and temperature for these alkanes by taking a random sample of conformers selected from the full ensemble. The probability  of a particular  conformer  being selected should be determined by the Boltzmann factor for that conformer, viz., $e^{-E_i / k_BT}$. To obtain all values of $E_i$ would require a calculation of the ground-state molecular energy for every conformer, which is also infeasible. However, changing the bond angle from \textit{trans} to \textit{gauche} requires an approximately constant amount of energy. The data in Table \ref{tab:n7conf} show that the value of $E_i$ is approximately proportional to the number of \textit{gauche} angles $N_g^{(i)}$: each \textit{gauche} angle contributes approximately 40--45~meV to the value of $E_i$. This is illustrated in Fig.~\ref{fig:heptane_DeltaE} for $n$-heptane, which shows that the dependence of $E_i$ on $N_g^{(i)}$ can be described by $E_i =40 N_g^{(i)}$ (solid black line).
We assume that this formula also gives a good approximation for the values of $E_i$ for larger $n$-alkanes ($n\geq8$)
and use it when generating conformers randomly, i.e., we replace the true Boltzmann factor $e^{-E_i / k_BT}$ by the estimated Boltzmann factor $\exp(- N_g^{(i)} \Delta E_{tg}/k_BT)$, with $\Delta E_{tg}=40$~meV. This estimate of the \textit{trans}-\textit{gauche} energy difference is based on the Hartree-Fock molecular energies. The results we present below can be adjusted to a different, e.g., smaller, value of $\Delta E_{tg}$, by rescaling the temperature axis.

The first step in generating a conformer randomly is to determine the number $N_g$ of \textit{gauche} angles it will have. This is governed by the probability distribution
\begin{align}
P(N_g) = \frac{\gamma(N_g) \exp(-N_g \Delta E_{tg} / k_BT) }{\sum_{N_g'} \gamma(N_g') \exp(-N_g' \Delta E_{tg} / k_BT)} , \label{eq:rand_prob_dist}
\end{align}
where $\gamma(N_g)$ is the number of possible conformers with $N_g$ \textit{gauche} angles (values in Table \ref{tab:no_conformers}), and the sum in the denominator runs from $N_g'=0$ to $N_g'=n-3$.
Once the value of $N_g$ has been determined, the next step is to determine how many of these \textit{gauche} angles should be $g^+$ and how many should be $g^-$. This is again decided randomly, with each \textit{gauche} angle having an equal chance of being $g^+$ or $g^-$.
The final step is to decide where each \textit{gauche} angle should be placed among the $n-3$ possible ``slots,'' with the remaining $n-3-N_g$ slots being designated as \textit{trans}. This is again determined randomly. If the end result is a conformer that has one or more $g^\pm g^\mp$ pairs (and thus is sterically strained), the conformer is discarded and we begin the process again from the first step.

Once a random sample of conformers has been generated, the positron binding energy $\varepsilon_b^{(i)}$ is calculated for each of them, and the average binding energy $\langle \varepsilon_b \rangle _T$ is estimated as the mean of these values. We have generated a random sample of 10 conformers for $n=7$, 10, 12, 14, and 16 at room temperature ($T=300$~K). For $n=7$, the results can be compared with those obtained using the full set of conformers (see below).
Table \ref{tab:conf_rand} lists the randomly generated conformers for each $n$, along with the corresponding positron binding energy (including the binding energy of the second bound state, where it exists).
\begin{table}
\caption{\label{tab:conf_rand}Randomly selected conformers of $n$-alkanes with $n=7$, 10, 12, 14, and 16 at room temperature ($T=300$~K), with their positron binding energies $\varepsilon_b^{(i)}$  for the first and (where applicable) second bound states.}
\begin{ruledtabular}
\begin{tabular}{ccccc}
&&& \multicolumn{2}{c}{$\varepsilon_b^{(i)}$ (meV)} \\
\cline{4-5}
$n$ & $N_g^{(i)}$ & Conformer $i$ & First & Second \\ 
\hline
7 & 1 & $tttg^-$ & 117.7 \\
   & 1 & $tttg^-$ & 117.7 \\
   & 3 & $tg^-g^-g^-$ & 132.1 \\
   & 1 & $tg^+tt$ & 118.4 \\
   & 2 & $tg^+g^+t$ & 127.4 \\
   & 1 & $tttg^+$ & 117.7 \\
   & 3 & $g^+ t g^- g^-$ & 129.7 \\
   & 1 & $g^- ttt$ & 117.7 \\
   & 1 & $tttg^-$ & 117.7 \\
   & 0 & $tttt$ & 111.5 \\
10 & 1 & $ttttg^-tt$ & 196.3 \\
     & 2 & $g^+tttttg^-$ & 195.7 \\
     & 1 & $ttg^+tttt$ & 196.3 \\
     & 2 & $tttg^-tttg^-$ & 203.1 \\ 
     & 3 & $tg^-tg^+g^+tt$ & 214.0 \\
     & 3 & $tg^-tg^+tg^-t$ & 202.2 \\
     & 1 & $ttttg^-tt$ & 196.3 \\
     & 2 & $g^-tg^-tttt$ & 206.0 \\
     & 0 & $ttttttt$ & 182.8 \\
     & 1 & $g^+tttttt$ & 189.4 \\
12 & 3 & $tg^+ttg^+ttg^+t$ & 252.1 \\   
     & 1 & $tttg^+ttttt$ & 232.9 & 0.1189 \\
     & 1 & $tttttttg^+t$ & 223.8 & 6.146 \\
     & 3 & $ttg^+ttg^-tg^+t$ & 249.8 \\
     & 4 & $g^+tg^+tg^+tttg^+$ & 256.5 & 0.1437 \\
     & 3 & $g^-ttg^+ttttg^-$ & 251.2 \\
     & 3 & $g^-ttg^-ttttg^+$ & 248.3 \\
     & 1 & $ttg^+tttttt$ & 230.0 & 1.944 \\
     & 2 & $g^+tg^+tttttt$ & 236.5 & 2.375 \\
     & 3 & $g^-ttttg^+tg^+t$ & 246.2 \\
14 & 3 & $g^-ttg^-tttttg^-t$ & 268.6 & 44.86 \\    
     & 1 & $g^+tttttttttt$ & 246.9 & 52.52 \\
     & 4 & $g^+ttttttg^+g^+tg^-$ & 269.2 & 47.39 \\
     & 2 & $g^+tg^+tttttttt$ & 270.1 & 45.82 \\
     & 3 & $tg^+tttttg^+ttg^-$ & 272.5 & 44.12 \\
     & 4 & $tttg^+ttg^+tg^-g^-t$ & 284.3 & 37.64 \\
     & 2 & $ttg^+ttg^-ttttt$ & 285.6 & 23.67 \\
     & 3 & $tttg^+ttttg^+tg^-$ & 267.6 & 43.35 \\
     & 2 & $ttg^+tttg^-tttt$ & 269.9 & 42.20 \\
     & 1 & $tttg^+ttttttt$ & 258.1 & 42.67 \\
16 & 3 & $tttg^-g^-ttttg^+ttt$ & 311.4 & 69.48 \\    
     & 1 & $tg^+ttttttttttt$ & 266.9 & 94.47 \\
     & 3 & $g^+tttttttttg^+tg^-$ & 275.5 & 99.08 \\
     & 3 & $tttttg^-ttg^-ttg^-t$ & 303.7 & 72.12 \\
     & 2 & $ttttttttg^+ttg^+t$ & 281.9 & 87.38 \\
     & 3 & $ttttg^-g^-ttttttg^-$ & 307.7 & 70.15 \\
     & 1 & $tttttttttg^-ttt$ & 275.7 & 88.62 \\
     & 5 & $tttg^+ttttg^-g^-tg^+g^+$ & 321.5 & 75.89 \\
     & 2 & $tttttttttg^-ttg^-$ & 279.0 & 91.87 \\
     & 3 & $tg^+ttttttg^+tttg^+$ & 299.4 & 83.62
\end{tabular}
\end{ruledtabular}
\end{table}
Note that in contrast to the calculations for $n\leq7$ using the full population of conformers, the molecular geometry has not been optimized at the Hartree-Fock  level using the 6--311++G$(d,p)$ basis. Rather, to reduce computational expense, the geometry has been optimized approximately using \textsc{avogadro}.\cite{avogadro} This results in smaller values of $\varepsilon_b^{(i)}$, but only by a few meV, e.g., the all-$t$ conformer for $n$-heptane has a positron binding energy of 117.2~meV using the fully optimized geometry (see Table \ref{tab:n7conf}) or 111.5~meV using the approximately optimized geometry (see Table \ref{tab:conf_rand}), a difference of 6~meV.

We obtain the following values of the average positron binding energy $\langle\varepsilon_b\rangle _T$ at room temperature ($T=300$~K): $n$-heptane, 120.8~meV; $n$-decane, 198.2~meV; $n$-dodecane, 242.7~meV; $n$-tetradecane, 269.3~meV; $n$-hexadecane, 292.3~meV. The binding energies of the corresponding all-$t$ conformers (using the approximately optimized geometry) are 111.5~meV, 182.8~meV, 216.7~meV, 242.1~meV, and 261.4~meV, respectively.
Thus, the average binding energy is 8.3\%, 8.4\%, 12\%, 11.2\%, and 11.8\% larger than that of the corresponding all-$t$ conformer, for $n=7$, 10, 12, 14, and 16, respectively. The value of $\langle \varepsilon_b \rangle _T$ obtained for $n$-heptane using the random sample (with approximately optimized geometry) is just 2.4~meV smaller than the value obtained using the full population (with fully optimized geometry). This difference is well within the uncertainty due to the different geometry optimization, and so we conclude that the random-sampling approach does provide reliable estimates of $\langle \varepsilon_b \rangle _T$.

A general trend that we observe for both small and large $n$-alkanes is that the positron binding energy tends to be larger for conformers that have a greater number of \textit{gauche} angles (see Tables \ref{tab:n7conf} and \ref{tab:conf_rand}). This effect is similar to that discussed in Sec.~\ref{sec:structure}, where we observed that
for $n>7$, the binding energies for cycloalkanes were greater than those for all-$t$ $n$-alkanes. Here, conformers with a greater number of \textit{gauche} angles are also more spatially compact, allowing the positron to sample the attractive molecular centres more effectively than for a fully stretched linear all-$t$ conformer.

\begin{figure}
\centering
\includegraphics[width=\columnwidth]{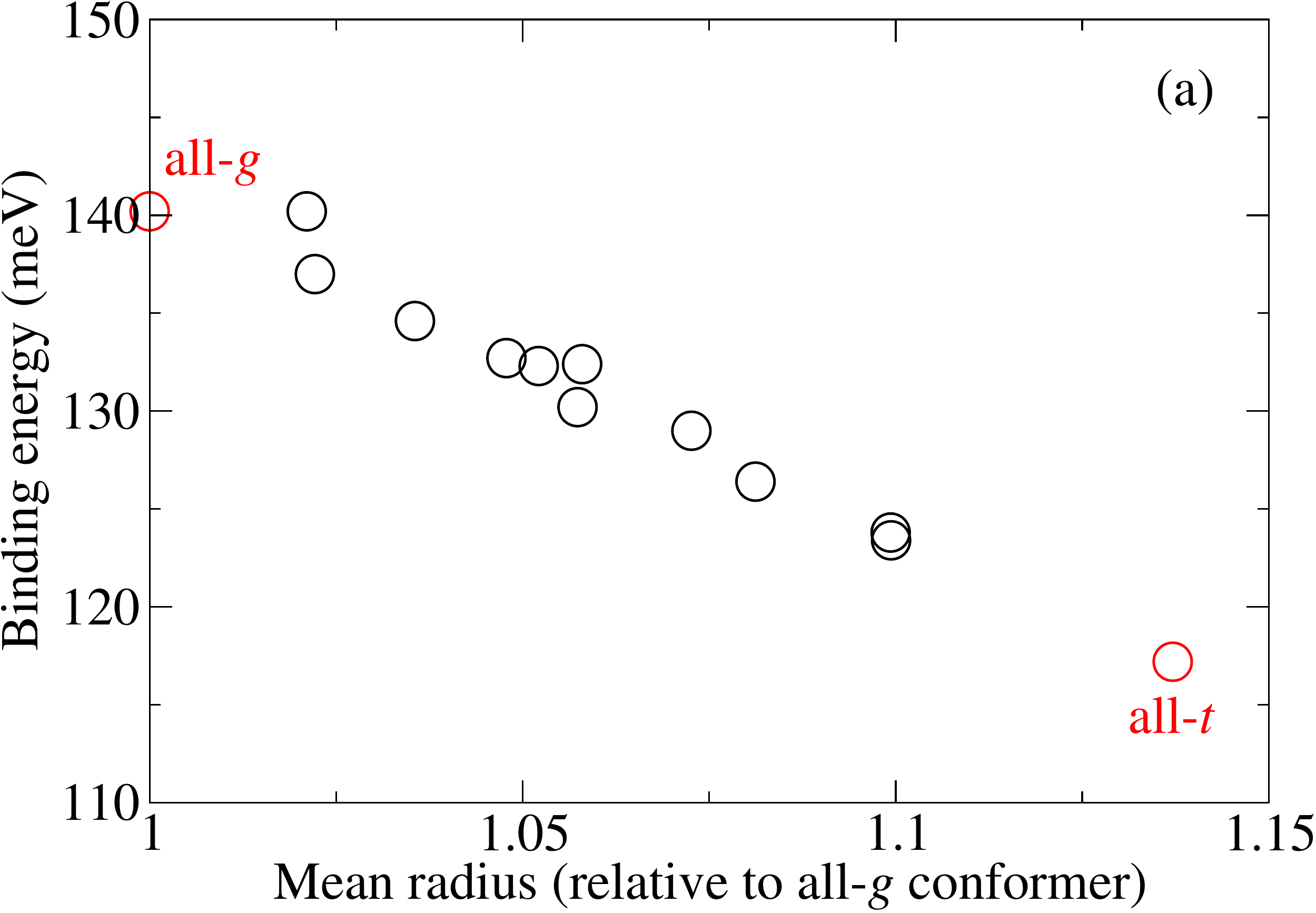}\\
\includegraphics[width=\columnwidth]{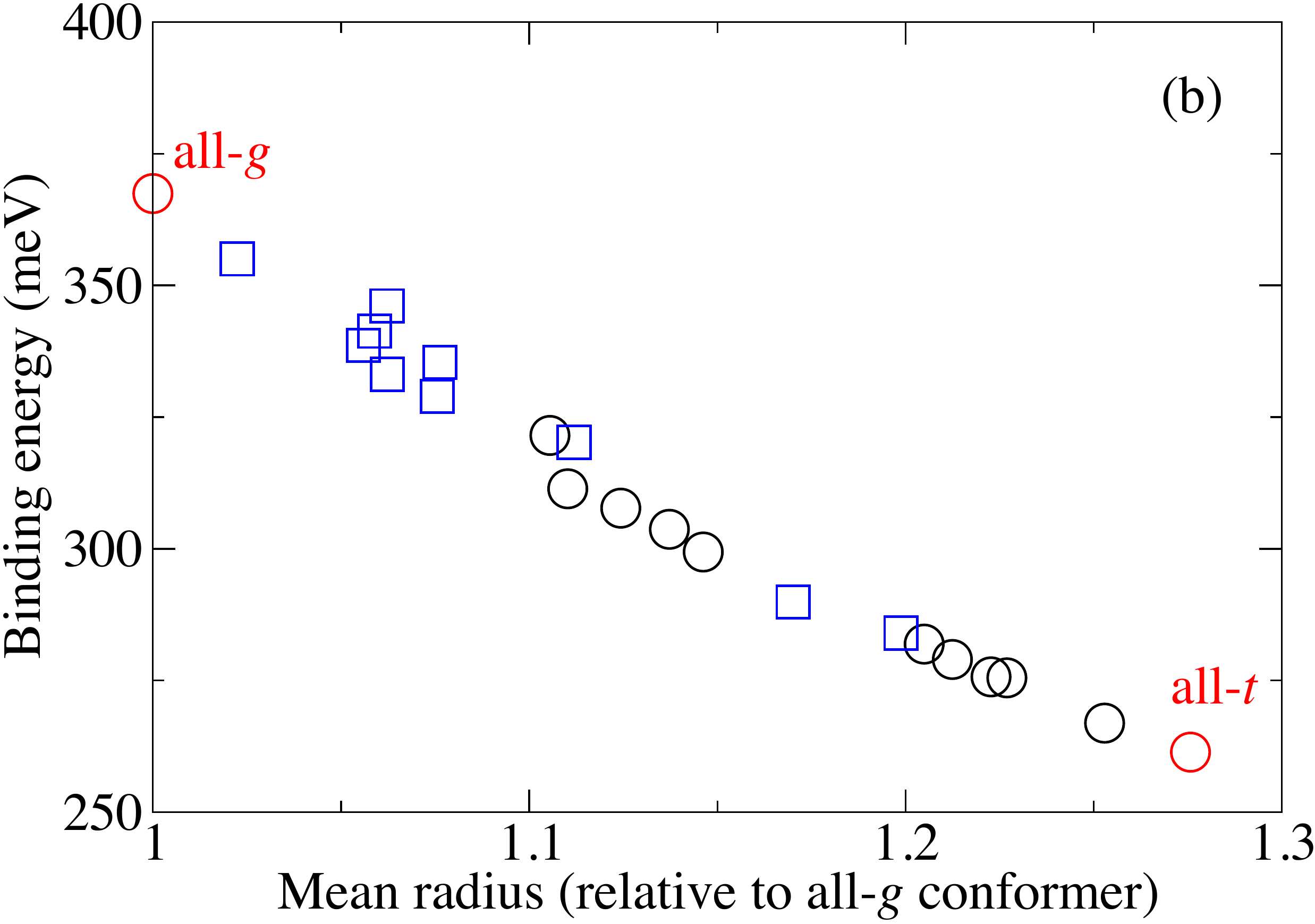}
\caption{\label{fig:compactness}Positron binding energies for the  conformers of (a) $n$-heptane C$_{7}$H$_{16}$ and (b) $n$-hexadecane C$_{16}$H$_{34}$, in terms of the mean radius of the conformer, relative to the  mean radius of the all-$g$ conformer. For heptane, the data for all spectroscopically different conformers are shown (Table~\ref{tab:n7conf}). In (b), black circles are for the conformers sampled at $T=300$~K (Table~\ref{tab:conf_rand}), while blue squares are those sampled in the $T\to \infty $ limit (Table~\ref{tab:Tinf_rand}).}
\end{figure}

To quantify the effect of the spatial extent of the conformer, we introduce the ``mean radius'' of the conformer,
\begin{align}
\langle r_A \rangle = \frac{1}{3n+2} \sum_{A=1}^{3n+2} \lvert \mathbf r_A - \mathbf r_\text{cm} \rvert ,
\end{align}
 where the sum is over all $3n+2$ atoms $A$ in the conformer, and $\mathbf r_\text{cm}$ is the position of the molecular center of mass.
Figure \ref{fig:compactness} shows the positron binding energy for $n$-heptane (full population of conformers) and $n$-hexadecane (random samples of conformers, along with the all-$g$ and all-$t$ conformers) as a function of the mean radius of the conformer, relative to the mean radius of the all-$g$ conformer.
In both cases, the binding energy increases almost linearly as the mean radius of the conformer decreases due to a larger number of \textit{gauche} angles.

\begin{figure}
\centering
\includegraphics[width=\columnwidth]{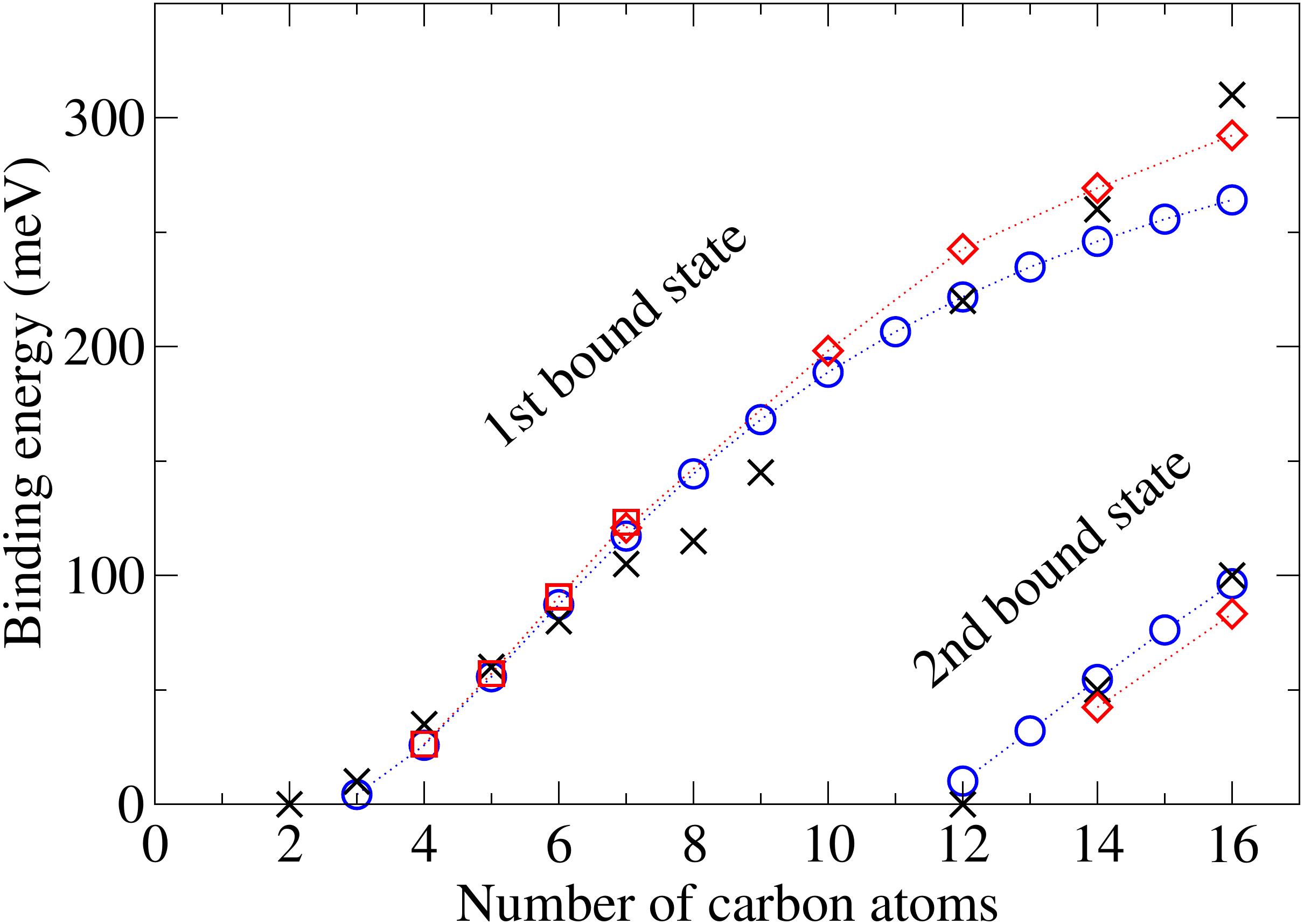}
\caption{\label{fig:conformers_average_eb}Positron binding energies for $n$-alkanes C$_n$H$_{2n+2}$. Blue circles, calculations for the all-$t$ conformer;\cite{Swann19} red squares, thermally averaged values calculated for all possible conformers; red diamonds, thermally averaged values calculated using a random sample of conformers; back crosses, experiment.\cite{Young08}}
\end{figure}

Figure \ref{fig:conformers_average_eb} shows the thermally averaged binding energies calculated using all the conformers for $n\leq 7$, and random sampling for $n\geq7$, as a function of $n$. Also shown are the original calculations for the all-$t$ conformers with fully optimized geometry,\cite{Swann19} and the experimental data.\cite{Young08}
For $n\lesssim7$, the thermally averaged binding energy is  close to the binding energy of the all-$t$ conformer. This is a consequence of the relatively weak conformer-dependence of the positron binding energy and small population of the conformers with \textit{gauche} angles.
For larger alkanes, the importance of considering single- and multiple-\textit{gauche} conformers becomes apparent. For $n=14$ and $n=16$, the calculated average binding energy is in better agreement with experiment than the binding energy of the all-$t$ conformer alone. At the same time, Fig.~\ref{fig:conformers_average_eb} shows that the thermally averaged binding energies still exhibit the same ``leveling off'' for sufficiently large values of $n$ that was observed for the all-$t$ conformers, while the experiment indicates that the binding energy continues to grow linearly with $n$. Note, however, that adopting a smaller value of the \textit{trans}-\textit{gauche} energy $\Delta E_{tg}$ would increase the thermally averaged  binding energies at $T=300$~K (see below). 

\begin{figure}
\centering
\includegraphics[width=0.95\columnwidth]{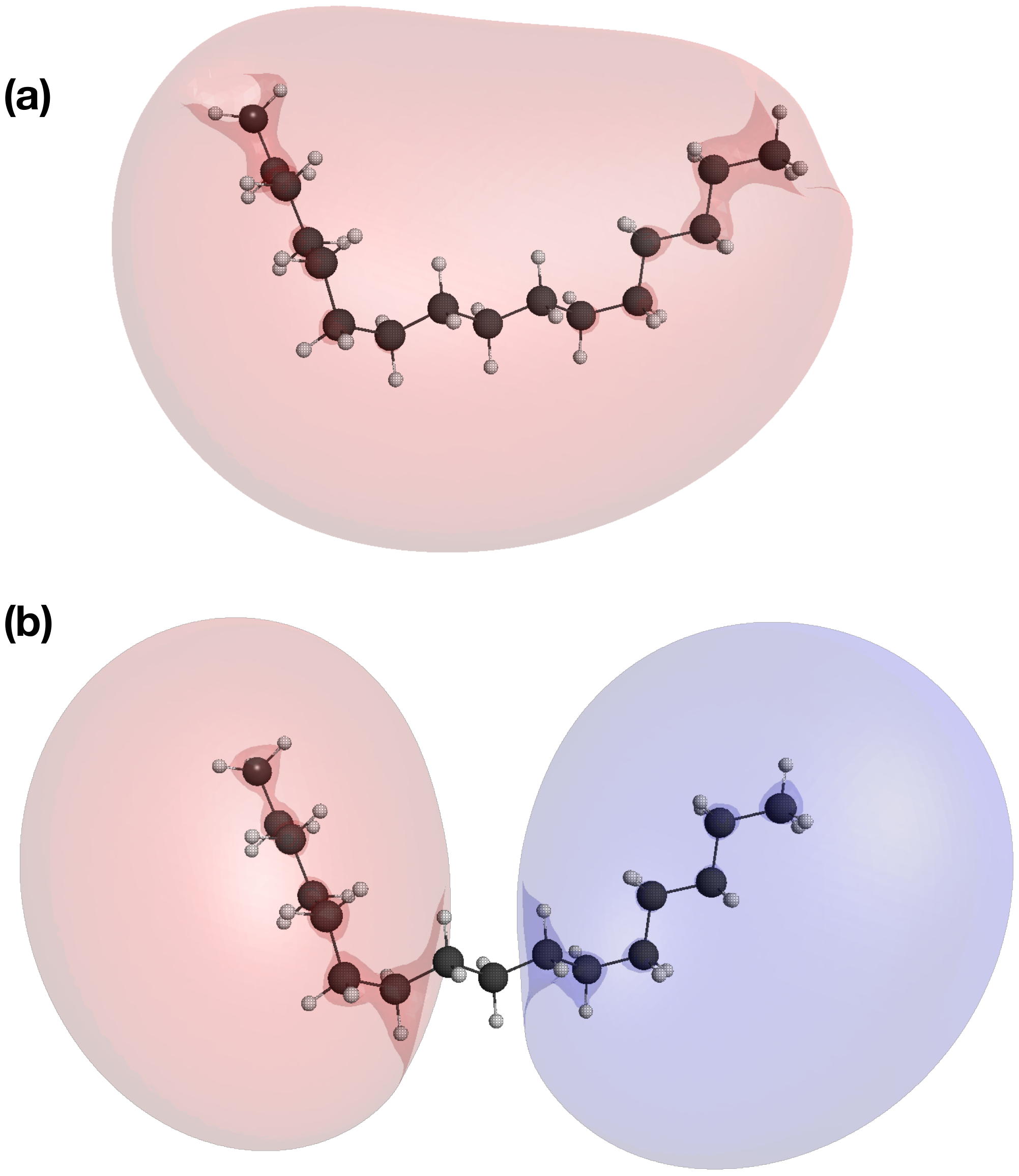} 
\caption{\label{fig:C16H34_surfaces}Wave function of the (a) first, and (b) second, bound positron state for the $tttg^-g^-ttttg^+ttt$ conformer of $n$-hexadecane C$_{16}$H$_{34}$. The wave function has a value of 0.00422~a.u. on the surfaces shown. Red (blue) indicates that the wave function is positive (negative).}
\end{figure}

Figure \ref{fig:conformers_average_eb}  also shows the average binding energies for the second bound states of $n$-tetradecane and $n$-hexadecane, which have values of  42.42~meV and 83.27~meV, respectively. Curiously, each of these values is smaller than the binding energy of the second bound state of the corresponding all-$t$ conformer (which has a value of 51.09~meV and 92.37~meV, respectively, using the approximately optimized geometry). In fact, the data in Table~\ref{tab:conf_rand} shows smaller second-state binding energies for conformers with larger numbers of \textit{gauche} angles. A simple explanation for this is as follows. 
The wave function of the first (ground) positron bound state surrounds the entire molecule and  has $s$-like character.
The wave function of the second positron bound state is orthogonal to that of the first bound state. Thus, it has a nodal surface near the center of the molecule where it changes sign and has a general $p$-like character (see Fig.~2 in Ref.~\onlinecite{Swann19}).
This means that the wave function of the  second bound state is ``more relaxed'' when the molecule is ``more linear.''
More precisely, the energy of the second bound state will be lower (i.e., the binding energy will be higher) for conformers that are more extended, i.e., have fewer \textit{gauche} angles, since the asymptotic $p$-like wave function is aligned along the near-straight axis of the molecule.
As an example, Fig.~\ref{fig:C16H34_surfaces} shows the wave functions of the first and second bound states for the $tttg^-g^-ttttg^+ttt$ conformer of $n$-hexadecane.
This particular conformer is U-shaped, and its mean  radius is 13\% smaller than that of the all-$t$ conformer.
The wave function of the first bound state embraces the entire molecule, the binding energy being 19\% greater than that of the all-$t$ conformer. On the other hand, the second bound state is more ``strained,'' as its two lobes are closer to each other than they would be in the case of the all-$t$ conformer. Its binding energy is 25\% smaller than that of the all-$t$ conformer.

\subsection{Temperature dependence of average binding energies}

\begin{figure*}
\centering
\includegraphics[width=0.8\textwidth]{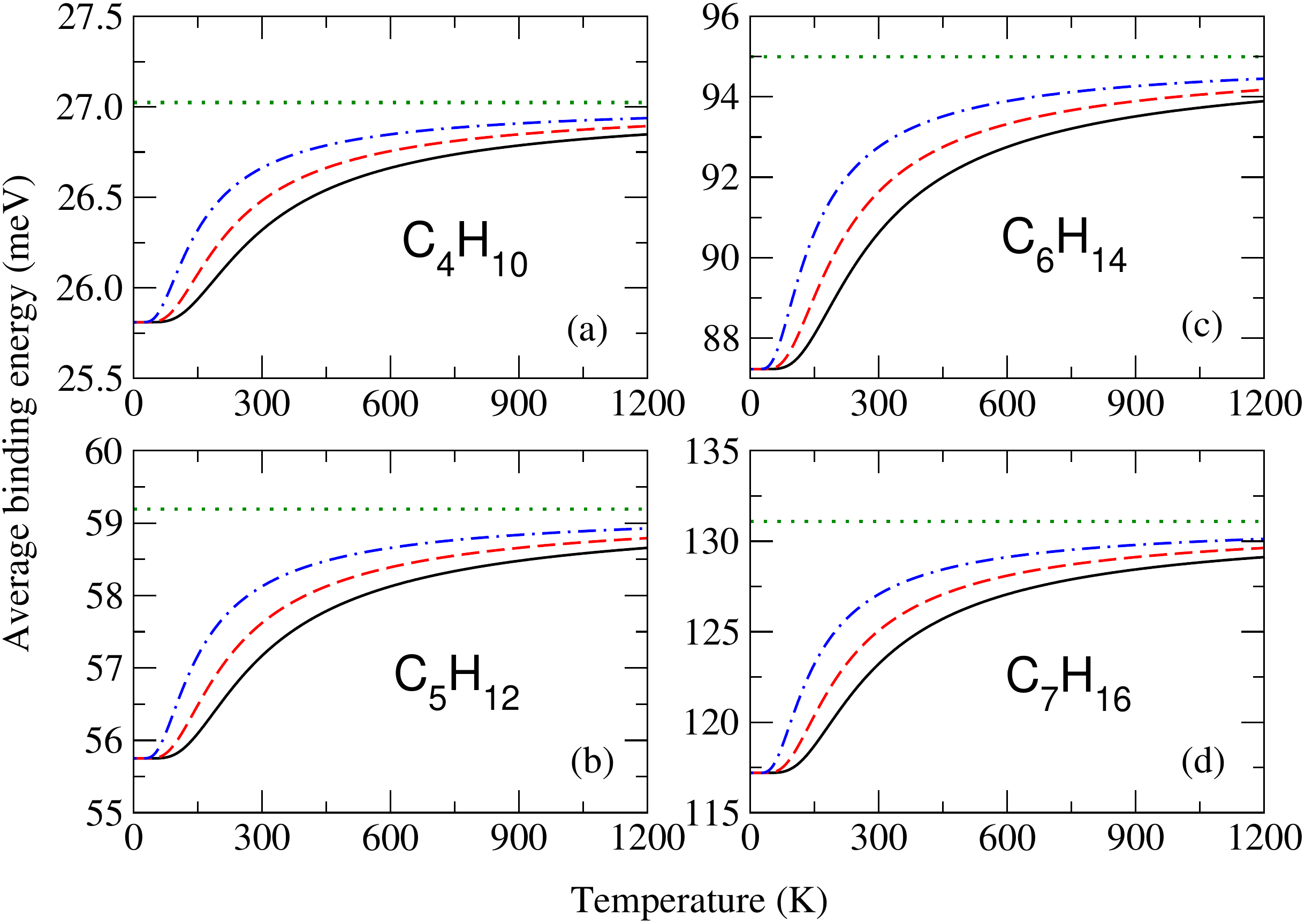}
\caption{\label{fig:n=4-7_full_T_dep}Average positron binding energy as a function of temperature for (a) $n$-butane C$_{4}$H$_{10}$, (b) $n$-pentane C$_{5}$H$_{12}$, (c) $n$-hexane C$_{6}$H$_{14}$, and (d) $n$-heptane C$_{7}$H$_{16}$, calculated for the full set of conformers using the Hartree-Fock energies (solid black curves), or those scaled by $\xi=0.75$ (dashed red curves) or $\xi=0.5$ (dot-dashed blue curves); dotted green lines show the asymptotic average binding energy for $T\to\infty$.}
\end{figure*}

So far, we have reported the calculations of conformer-averaged binding energies for room temperature, $T=300$~K. We now investigate the temperature dependence of the average binding energies.
For $n\leq 7$, where we have calculated binding energies for all of the conformers, this can be done directly using Eqs.~(\ref{eq:prob-dist}) and (\ref{eq:av_bine}).
Figure \ref{fig:n=4-7_full_T_dep} shows the average binding energy for butane through to heptane as a function of temperature, for $T\leq1200$~K (solid black curves).
In the limit of low $T$, all of the Boltzmann factors $e^{-E_i/k_BT}$  in Eq.~(\ref{eq:prob-dist})  tend to zero, with the exception of that of the lowest-energy (all-$t$) conformer, for which $E_i=0$. Thus, as $T\to0$ the average binding energy approaches the binding energy of the all-$t$ conformer. In the limit of large $T$, all of the Boltzmann factors tend to unity. Thus, as $T\to\infty$, all possible conformers contribute equally to the average binding energy (weighted by their degeneracies $w_i$). The asymptotic large-$T$ limit is indicated by a dotted green line in Fig.~\ref{fig:n=4-7_full_T_dep}.

The values of $E_i$ used for $n=4$--7 were obtained from molecular ground-state energy calculation in the Hartree-Fock approximation. However, it is likely that these values are larger than the true values. Post-Hartree-Fock calculations indicate that the physical values of $E_i$ may be as much as a factor of 2 smaller than the Hartree-Fock values (see, e.g., Refs.~\onlinecite{Smith96,Balabin08,Csontos16}). Thus, we also show in Fig.~\ref{fig:n=4-7_full_T_dep} how the temperature dependence of $\langle \varepsilon_b \rangle _T$ changes if we scale the Hartree-Fock energies $E_i$ by a factor of $\xi=0.75$  or $\xi=0.5$ ($\xi=1$ denotes the unscaled Hartree-Fock values). 
Of course, such scaling of the energies is equivalent to increasing the temperature by a factor of $1/\xi$. As expected, the reduced values of $E_i$ yield  larger values of $\langle\varepsilon_b\rangle _T$ for all $T>0$, as a result of increasing populations of
conformers with single- and multiple-\textit{gauche} angles.

For the larger $n$-alkanes, we have used probabilistic sampling to estimate the room-temperature value of $\langle\varepsilon_b\rangle _T$, so its temperature dependence cannot be determined directly. We therefore estimate it according to the following procedure. For each $n$-alkane, in addition to the existing random sample of 10 conformers taken at $T= T_0=300$~K with $\Delta E_{tg}=40$~meV, we take another random sample of 10 conformers for $T\to\infty$. In this limit, the probability for a conformer to have $N_g$ \textit{gauche} angles is
\begin{align}
P(N_g) = \frac{\gamma(N_g)}{\sum_{N_g'} \gamma(N_g') } ; \label{eq:rand_prob_dist_Tinf}
\end{align}
cf. Eq.~(\ref{eq:rand_prob_dist}). We use this sample to estimate the value of $\langle\varepsilon_b\rangle_T$ for $T\to\infty$. 
Table \ref{tab:Tinf_rand} shows the random samples for $T\to\infty$ for $n=7$, 10, 12, 14, and 16. 
\begin{table}
\caption{\label{tab:Tinf_rand}Conformers of $n$-alkanes with $n=7$, 10, 12, 14, and 16, selected randomly for $T\to\infty$, with their positron binding energies $\varepsilon_b^{(i)}$ for the first and (where applicable) second bound states.}
\begin{ruledtabular}
\begin{tabular}{ccccc}
&&& \multicolumn{2}{c}{$\varepsilon_b^{(i)}$ (meV)} \\
\cline{4-5}
$n$ & $N_g^{(i)}$ & Conformer $i$ & First & Second \\ 
\hline
7 & 3 & $g^+g^+tg^-$ & 129.8 \\
   & 3 & $tg^-g^-g^-$ & 132.2 \\
   & 2 & $g^+ttg^-$ & 126.8 \\
   & 3 & $g^+g^+g^+t$ & 132.2 \\
   & 3 & $g^+tg^-g^-$ & 129.8 \\
   & 2 & $g^-ttg^-$ & 124.8 \\
   & 1 & $tg^+tt$ & 118.4 \\
   & 3 & $tg^-g^-g^-$ & 132.2 \\
   & 3 & $g^-g^-tg^-$ & 135.5 \\
   & 1 & $tttg^-$ & 117.8 \\
10 & 2 & $g^-ttg^+ttt$ & 208.8 \\
     & 2 & $g^-ttg^-ttt$ & 204.4 \\ 
     & 4 & $g^-g^-tg^+ttg^-$ & 219.2 \\
     & 2 & $ttttg^-tg^-$ & 202.1 \\
     & 4 & $tg^-g^-tg^+tg^+$ & 221.6 \\
     & 4 & $tg^-g^-tg^-tg^+$ & 228.9 \\
     & 4 & $tg^+tg^-g^-tg^+$ & 218.4 \\
     & 3 & $tg^-g^-ttg^+t$ & 227.6 \\
     & 5 & $ttg^+g^+g^+g^+g^+$ & 231.2 \\
     & 3 & $tg^+g^+tg^-tt$ & 213.2 \\
12 & 4 & $g^+tg^+ttg^-g^-tt$ & 284.5 \\    
     & 6 & $tg^+g^+tg^+g^+tg^-g^-$ & 315.7 \\
     & 3 & $ttttg^+g^+ttg^-$ & 264.4 \\
     & 6 & $tg^-g^-g^-tg^+tg^-g^-$ & 274.7 \\
     & 6 & $g^+g^+tg^+tg^+tg^+g^+$ & 278.1 \\
     & 5 & $g^+g^+g^+tg^+ttg^-t$ & 291.2 \\
     & 5 & $tg^+g^+tg^+g^+tg^+t$ & 334.0 \\
     & 4 & $g^+g^+g^+ttttg^+t$ & 252.9 \\
     & 5 & $g^-tttg^-g^-g^-tg^+$ & 277.7 \\
     & 7 & $g^+g^+g^+g^+tg^-g^-tg^+$ & 281.8 \\
14 & 2 & $tttg^+ttttttg^+$ & 261.0 & 45.33 \\     
     & 4 & $tg^-tg^+ttg^+ttg^-t$ & 293.0 & 27.59 \\
     & 7 & $tg^+g^+tg^+tg^-tg^+g^+g^+$ & 307.6 & 38.29 \\
     & 7 & $tttg^-g^-g^-tg^-g^-g^-g^-$ & 370.9 & 66.89 \\
     & 5 & $tg^+tg^-g^-tg^-tttg^-$ & 316.1 & 24.48 \\
     & 5 & $g^-ttg^+g^+tg^+g^+ttt$ & 436.0 \\
     & 6 & $tg^-g^-ttg^-tg^-g^-g^-t$ & 345.2 & 92.21 \\
     & 7 & $g^+tg^+tg^-g^-tg^-g^-tg^-$ & 380.6 & 59.97 \\
     & 9 & $g^-g^-g^-g^-tg^+g^+tg^-g^-g^-$ & 323.0 & 38.93 \\
     & 5 & $g^+tg^-g^-tg^-tg^-ttt$ & 307.7 & 37.06 \\
16 & 6 & $g^-tg^+g^+tttg^+ttg^+tg^+$ & 333.1 & 71.34 \\     
     & 8 & $g^-tg^+g^+tg^-g^-tg^+ttg^-g^-$ & 341.3 & 91.64 \\
     & 7 & $g^+ttg^+tg^-tg^-tg^+g^+tg^-$ & 320.2 & 93.05 \\
     & 8 & $g^-ttg^+tg^-g^-tg^+g^+tg^-g^-$ & 335.4 & 93.28 \\
     & 8 & $g^+tg^-g^-tg^-tg^+g^+ttg^+g^+$ & 355.0 & 73.70 \\
     & 6 & $g^-tg^-tg^-ttg^-tttg^-g^-$ & 328.9 & 74.39 \\
     & 3 & $tg^+ttg^+ttttttg^+t$ & 289.9 & 87.58 \\
     & 3 & $tg^+tg^+tttttttg^+t$ & 284.0 & 95.68 \\
     & 7 & $g^+tttg^+tg^+ttg^-g^-g^-g^-$ & 346.1 & 81.49 \\
     & 5 & $g^+tttg^-ttg^-tg^-ttg^+$ & 338.6 & 62.21
\end{tabular}
\end{ruledtabular}
\end{table}
Next, we approximate $\langle\varepsilon_b\rangle_T$ by
\begin{align}
\langle\varepsilon_b\rangle _T = \langle\varepsilon_b\rangle _0 + [\langle\varepsilon_b\rangle _\infty - \langle\varepsilon_b\rangle _0] e^{-\tau/T}, \label{eq:T_dep_rand}
\end{align}
where $\tau$ is a parameter. Requiring that Eq.~(\ref{eq:T_dep_rand})  give the same value for $\langle\varepsilon_b\rangle _{T_0}$ as the original random sample for $T=T_0$ provides the value of $\tau$, viz.,
\begin{align}
\tau = T_0 \ln \frac{\langle\varepsilon_b\rangle _\infty -\langle\varepsilon_b\rangle _0}{\langle\varepsilon_b\rangle _{T_0}-\langle\varepsilon_b\rangle _0} .
\end{align}

\begin{figure}
\centering
\includegraphics[width=\columnwidth]{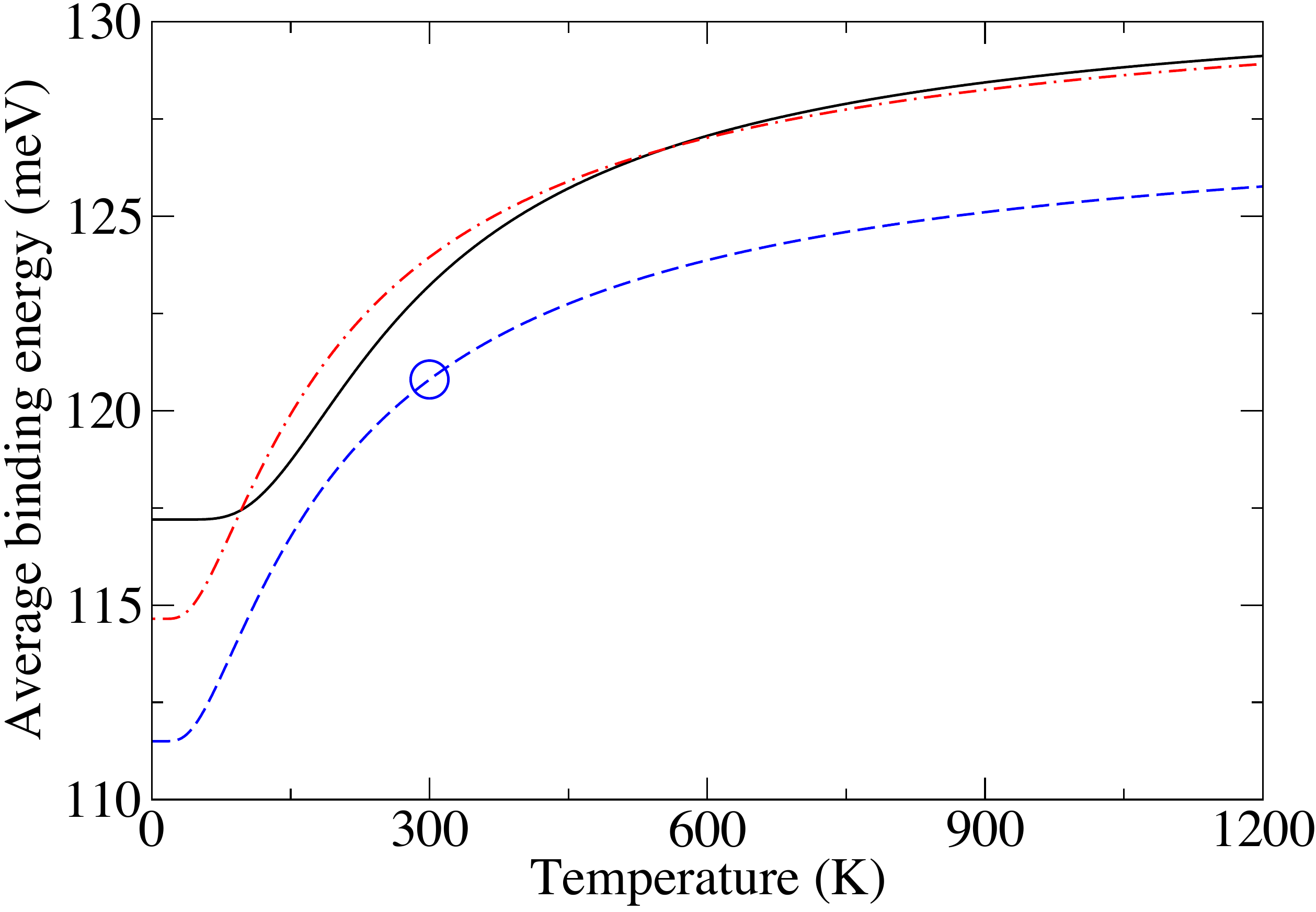}
\caption{\label{fig:heptane_T_dep}Average positron binding energy for $n$-heptane C$_7$H$_{16}$ as a function of temperature. Solid black curve, calculation using full set of conformers with fully optimized geometry; blue circle, calculation for $T=300$~K using random sample with $\Delta E_{tg}=40$~meV and  approximately optimized geometry; dashed blue curve, Eq.~(\ref{eq:T_dep_rand}) with $\tau=171.1$~K; dot-dashed red curve, same as dashed blue curve but shifted upwards to match the black curve at $T\to \infty $.}
\end{figure}

Figure~\ref{fig:heptane_T_dep} shows $\langle\varepsilon_b\rangle _T$ for $n$-heptane,
obtained using all of the conformers with the fully optimized geometry and $\xi=1$ (solid black curve), and the data obtained using random samples with the approximately optimized geometry (dashed blue curve); the value of $\tau$ is 171.1~K. Qualitatively, the two methods predict a similar temperature dependence of $\langle\varepsilon_b\rangle_T$; the difference between them is predominantly due to the different optimization of the molecular geometry. Indeed, shifting the curve obtained using random samples (with approximately optimized geometry) upwards so that its asymptotic $T\to\infty$ limit matches that obtained using the full population (with fully optimized geometry) gives a much better agreement with the full-population curve across the range of temperatures shown (the result is shown by the dot-dashed red curve).

\begin{figure*}
\centering
\includegraphics[width=0.8\textwidth]{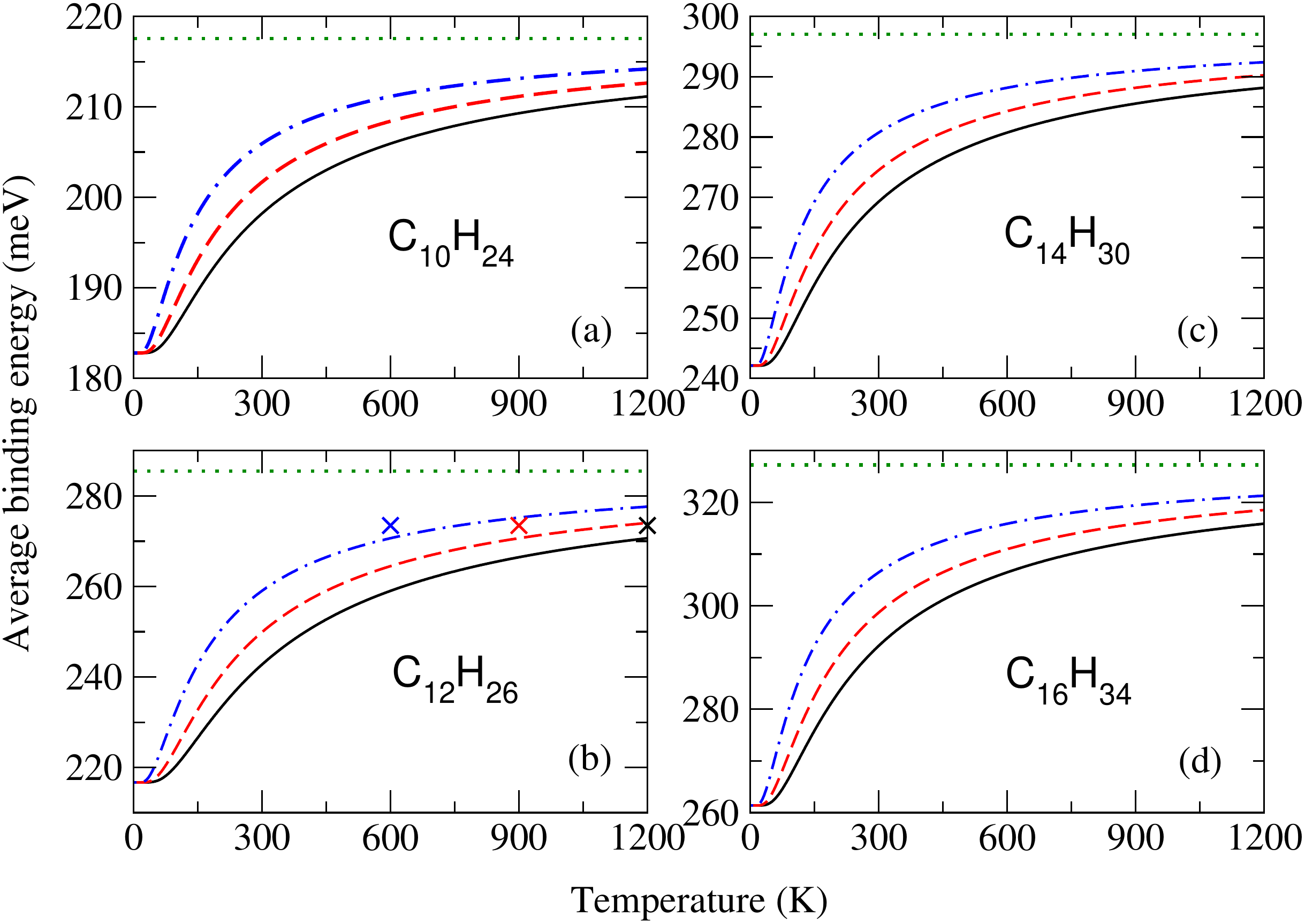}
\caption{\label{fig:n=10-16_random_T_dep}Average positron binding energy as a function of temperature for (a) $n$-decane C$_{10}$H$_{22}$, (b) $n$-dodecane C$_{12}$H$_{26}$, (c) $n$-tetradecane C$_{14}$H$_{30}$, and (d) $n$-hexadecane C$_{16}$H$_{34}$, calculated using random samples of conformers. Solid black curves, $\Delta E_{tg}=40$~meV; dashed red curves, $\Delta E_{tg}=30$~meV; dot-dashed blue curves, $\Delta E_{tg}=20$~meV; dotted green lines, asymptotic average binding energy as $T\to\infty$. Crosses in (b): random sample with $\Delta E_{tg}=40$~meV, $T=1200$~K (black); $\Delta E_{tg}=30$~meV, $T=900$~K (red); $\Delta E_{tg}=20$~meV, $T=600$~K (blue).}
\end{figure*}

Figure ~\ref{fig:n=10-16_random_T_dep} shows the predicted temperature dependence of $\langle\varepsilon_b\rangle _T$ obtained from Eq.~(\ref{eq:T_dep_rand}) for $n=10$, 12, 14, and 16.
Besides the result for $\Delta E_{tg}=40$~meV, the figure also shows $\langle\varepsilon_b\rangle _T$ for the smaller values of $\Delta E_{tg}=30$~meV  and 20~meV, i.e., the results of scaling the original value of $\Delta E_{tg}$ by a factor of  $\xi=0.75$ or 0.5. These are obtained simply by taking the curve for $\Delta E_{tg}=40$~meV and scaling the parameter $\tau$ by $\xi$.
To verify that the temperature dependence of $\langle\varepsilon_b\rangle _T$ estimated in this manner is reliable across the range of temperatures, we have calculated $\langle\varepsilon_b\rangle _T$ for $n$-dodecane at $T=1200$~K using an additional random sample of 10 conformers with $\Delta E_{tg}=40$~meV. The result is shown in Fig.~\ref{fig:n=10-16_random_T_dep}(b) (black cross). Also shown are the equivalent values of $\langle\varepsilon_b\rangle _T$ at $T=900$~K with $\Delta E_{tg}=30$~meV  (red cross), or at $T=600$~K with $\Delta E_{tg}=20$~meV (blue cross), corresponding to the same ratio $\Delta E_{tg}/k_BT = 0.39$.
In each case, the value of $\langle\varepsilon_b\rangle _T$ predicted using Eq.~(\ref{eq:T_dep_rand}) at the specified value of $T$ is very close to the value obtained from the random sample at this value of $T$.

Figures \ref{fig:n=4-7_full_T_dep} and \ref{fig:n=10-16_random_T_dep} show that values of $\langle\varepsilon_b\rangle _T$ show high sensitivity to the molecular temperature for $T\sim 300$~K. This suggests that positron annihilation studies can be used as a ``molecular thermometer,'' or as probe of the conformational distribution of molecular ensembles, or as a means for determining the value of the \textit{trans}-\textit{gauche} energy difference. 

\subsection{Resonant annihilation rate}

In this section we analyze how the presence of conformers would affect the positron annihilation rates, as measured with a trap-based positron beam.\cite{Gilbert97,Gilbert02,Gribakin10,Natisin16}
The positron annihilation rate $\lambda$ in a gas of number density $n$ is related to the annihilation cross section $\sigma_a$ by
\begin{align}
\lambda = \sigma_a v n ,
\end{align}
where $v$ is the positron velocity. The annihilation cross section is traditionally expressed in terms of the dimensionless normalized annihilation rate $\Zeff$,
\begin{align}
\sigma_a = \pi r_0^2 \frac{c}{v} \Zeff , \label{eq:ann_cs}
\end{align}
where $r_0$ is the classical electron radius, and $c$ is the speed of light. The quantity $\Zeff$ is referred to as the effective number of electrons that contribute to the positron annihilation for a given target atom or molecule.\footnote{Setting $\Zeff =1$ recovers the basic Dirac cross section of electron-positron annihilation in the nonrelativistic approximation.}
Naturally, both $\sigma _a$ and $\Zeff $ are functions of the incident positron energy $\varepsilon $.

As mentioned in Sec.~\ref{sec:intro}, for most molecules, the measured positron-molecule annihilation rate displays prominent peaks at specific positron energies.
According to the current understanding of positron annhilation with molecules, there are two mechanisms for annihilation to proceed, viz., \textit{direct} and \textit{resonant}.\cite{Gribakin00,Gribakin01,Gribakin10} Direct annihilation occurs when the incident positron annihilates with a target electron ``in flight.'' Its contribution to the total annihilation rate can be enhanced at low energies if the molecule has a low-lying virtual level or weakly bound state for the positron. However, the possible $\Zeff$ values due to such effect are limited to $\Zeff <10^3$ for low-energy (room-temperature) positron energies. Resonant annihilation can only occur for molecules that have a bound state for the positron. In this case the positron is captured by the molecule, forming a VFR, from which the positron can annihilate with an electron (or it can detach from the molecule again). Resonant annihilation requires  the energy  of the incident positron to be close to the energy of a VFR channel, $\varepsilon _\nu =\hbar\omega_\nu-\varepsilon_b>0$, where $\hbar \omega_\nu$ is the vibrational excitation energy of the molecule and $\varepsilon_b$ is the positron binding energy. The  peaks in the  annihilation rate occur at the energies of the VFRs, and their contribution far exceeds that of direct annihilation for all but the simplest molecules.\cite{Gilbert02,Barnes03,Barnes06,Young08,Young08a,Gribakin10}

The resonant part of the annihilation cross section is given by the Breit-Wigner formula
(in atomic units, a.u.):\cite{LandauQM}
\begin{align}
\sigma_a(\varepsilon) = \frac{\pi}{k^2} \sum_\nu \frac{\Gamma^a_\nu \Gamma^e_\nu}{(\varepsilon - \omega_\nu + \varepsilon_b)^2 + \Gamma_\nu^2/4} , \label{eq:ann_cs_BW}
\end{align}
where $\varepsilon$ is the incident positron energy; $k=\sqrt{2\varepsilon}$ is the positron momentum; $\Gamma^a_\nu$, $\Gamma^e_\nu$, and $\Gamma_\nu$ are the   annihilation, elastic, and total width of the $\nu$th resonance, respectively; and the sum is over all resonances. Comparing Eqs.~(\ref{eq:ann_cs}) and (\ref{eq:ann_cs_BW}) gives the following expression for the resonant part of $\Zeff$:
\begin{align}
\Zeff(\varepsilon) = \frac{\pi}{k} \delta_{ep} \sum_\nu \frac{\Gamma^e_\nu}{(\varepsilon - \omega_\nu + \varepsilon_b)^2 + \Gamma_\nu^2/4} , \label{eq:Zeff_BW}
\end{align}
where we have  used the fact that the annihilation width is proportional to the average electron density at the positron (or contact density) in the bound state, $\delta_{ep}$:\cite{Gribakin00,Gribakin01}
\begin{align}
\Gamma^a_\nu = \pi r_0^2 c \delta_{ep} .
\end{align}

Equation (\ref{eq:Zeff_BW}) can be used to calculate $\Zeff$ as a function of the incident positron energy $\varepsilon$. We wish to determine what effect the presence of multiple conformers in the gas has on the resonances in $\Zeff$. Since each conformer has a different binding energy $\varepsilon_b$ (and, strictly speaking, slightly different vibrational energies), the observed peaks in the $\Zeff$ spectrum will, in fact, be averages of several peaks due to the different conformers. Thus, we expect the observed resonances to be broader than those for a single conformer.

To enable a comparison with experiment, Eq.~(\ref{eq:Zeff_BW}) should be averaged over the energy distribution of the positron beam, which can be modeled by the combination of a Gaussian distribution in the longitudinal direction ($z$) and a Maxwellian distribution in the transversal direction $(\perp$),\cite{Gribakin06,Gribakin10} 
\begin{align}
f(\varepsilon_\perp , \varepsilon_z) = \frac{1}{k_B T_\perp \sqrt{2\pi\sigma_z^2}} \exp \left[ -\frac{\varepsilon_\perp}{k_BT_\perp} - \frac{(\varepsilon_z - \epsilon)^2}{2\sigma_z^2} \right] ,
\end{align}
where
$\varepsilon_\perp$ and $\varepsilon_z$ are the transverse and longitudinal positron energies ($\varepsilon=\varepsilon_\perp+\varepsilon_z$),
 $T_\perp$ is the effective transverse temperature of the beam, $\epsilon$ is the mean longitudinal (or parallel) energy of the positrons (as measured by the retarding potential analyzer), and $\sigma_z$ is the root-mean-squared
width of the parallel energy distribution ($\sigma_z=\delta_z/\sqrt{8\ln2}$, with $\delta_z$ being the full width at half maximum).
$\Zeff$ is averaged over the energy distribution of the beam according to
\begin{align}
\overline Z_\text{eff}(\epsilon) = \iint Z_\text{eff}(\varepsilon) f(\varepsilon_\perp , \varepsilon_z) \, d\varepsilon_\perp \, d\varepsilon_z . \label{eq:zeff_bar}
\end{align}
Since the resonance widths $\Gamma_\nu$ are small compared to the energy spread of the beam, 
%and the total width of each resonance is dominated by its elastic width (i.e., $\Gamma_\nu\approx\Gamma^e_\nu$),\cite{Gribakin06}
 the Breit-Wigner profiles in Eq.~(\ref{eq:Zeff_BW}) can be replaced by $\delta$ functions:\cite{Gribakin06}
\begin{align}
\Zeff = \frac{2\pi^2}{k} \delta_{ep} \sum_\nu \frac{\Gamma^e_\nu}{\Gamma_\nu} \delta(\varepsilon-\omega_\nu+\varepsilon_b). 
\end{align}
 Equation (\ref{eq:zeff_bar}) then gives
\begin{align}
\overline Z_\text{eff}(\epsilon) = 2\pi^2 \delta_{ep} \sum_\nu \frac{\Gamma^e_\nu}{k_\nu \Gamma_\nu} \Delta(\epsilon-\omega_\nu+\varepsilon_b) , \label{eq:zeff_bar_final}
\end{align}
where $k_\nu=[2(\omega_\nu-\varepsilon_b)]^{1/2}$, and
\begin{align}
\Delta(E) &= \frac{e^{\sigma_z^2/2(k_BT_\perp)^2}}{2k_BT_\perp} e^{E/k_BT_\perp} 
\left\{ 1+ \operatorname{erf} \left[ -\frac{1}{\sqrt 2} \left( \frac{E}{\sigma_z} + \frac{\sigma_z}{k_B T_\perp} \right) \right] \right\} . \label{eq:Delta}
\end{align}
The function $\Delta(\epsilon-\omega_\nu+\varepsilon_b)$ is asymmetric, with a low-energy ``tail'' due to the positron transverse energy content that allows it to access a resonance for $\epsilon<\omega_\nu-\varepsilon_b$. Thus, the maxima of the resonant peaks described by Eq.~(\ref{eq:zeff_bar_final}) are downshifted from the resonance energies $\varepsilon_\nu=\omega_\nu-\varepsilon_b$.
For a typical positron buffer-gas trap (BGT), $k_BT_\perp=20$~meV and $\sigma_z=10$~meV; for a cryogenic beam-tailoring trap (CBT), $k_BT_\perp=5$~meV and $\sigma_z=8$~meV.\footnote{J. R. Danielson, S. Ghosh, and C. M. Surko (private communication).}

In principle, the sum in Eq.~(\ref{eq:zeff_bar_final}) should be over all vibrational excitations of the molecule with $\omega_\nu>\varepsilon_b$, both single-quantum (i.e., fundamentals) and multi-quantum (i.e., overtones and combinations). However, from early on, the experimental data for energy-resolved resonant $\Zeff$ showed that their energy dependence for a given molecule  is similar to the spectrum of the fundamentals alone.\cite{Marler04,Barnes06,Gribakin10} For small polyatomics (e.g., methyl halides), the VFRs of the fundamentals provide an accurate description of their $\Zeff$.\cite{Gribakin06} For larger polyatomics, however, the magnitudes of the peaks in $\Zeff$ far exceed the contribution of the single-quantum VFRs. In this case, the simple, fundamental vibrations act as \textit{doorways} into the dense spectrum of multimode vibrations,\cite{Gribakin04,Gribakin10,Gribakin17} in a process similar to intramolecular vibrational energy redistribution (IVR). As a result, the energy dependence of $\Zeff$ retains a strong resemblance with the spectrum of the fundamentals. Equation~(\ref{eq:zeff_bar_final}) can then be used for modeling or fitting experimental $\Zeff$ data, by restricting the sum to the normal modes only and considering the ratios $\Gamma^e_\nu /\Gamma_\nu $ as free parameters.\cite{Gribakin09,Danielson13,Natisin:thesis}

Here we will assume for simplicity that the ratio $\Gamma^e_\nu/\Gamma_\nu \equiv \beta _\nu $ has  approximately  the same  value ($\beta $) for all vibrational modes $\nu$. To justify this, we first recall that the annihilation width $\Gamma^a_\nu$ is expected to be much smaller than the elastic widths $\Gamma^e_\nu$ of the fundamentals, so $\Gamma_\nu = \Gamma^a_\nu+\Gamma^e_\nu \approx \Gamma^e_\nu$ and $\beta \approx 1$ in small molecules.\cite{Gribakin06} In large molecules, the fundamentals play the role of doorways, and $\Gamma^e_\nu$ can be taken as the elastic widths of the single-quantum, doorway excitation, while the total width $\Gamma _\nu$ in the denominator is that of the ultimate multimode vibrational states.\cite{Gribakin04} The latter should have little level-to-level variation, as it contains contributions from many positron escape channels.\cite{Gribakin09} The elastic width, on the other hand, can vary vary significantly between the fundamentals, and so will $\beta _\nu$. When modelling the annihilation rate in the energy range of the C-H-stretch peak for large alkanes, one includes contributions of a large number ($2n+2$) of C-H-stretch fundamentals, so the value of $\beta $ can be taken as the average for this group. Here one must also note that first-principle calculations of these factors for alkane molecules with more than a few carbons are beyond the capability of present theoretical approaches.\cite{Gribakin17}

For weakly bound states, the $\delta_{ep}$ factor in Eq. (\ref{eq:zeff_bar_final}) is proportional to the square root of the positron binding energy, viz.,\cite{Gribakin01}
\begin{align}
\delta_{ep}=\frac{F}{2\pi}\kappa , \label{eq:deltaep}
\end{align}
where $F$ is a constant and $\kappa=\sqrt{2\varepsilon_b}$ (see also Ref. \onlinecite{Mitroy02}). This dependence is also confirmed in the calculations of positron bound states with alkanes.\cite{Swann19} Analysis of high-quality configuration-interaction and stochastic-variational calculations of the annihilation rates in positron-atom bound states gave a value of $F=0.66$~a.u. for atoms,\cite{Gribakin01} although the value may be slightly larger for molecules.\cite{Swann19} 
Our chief interest is in the effect of the presence of multiple conformers on the shape and positions of the resonances in $\overline Z_\text{eff}$, and not on the overall magnitude of $\overline Z_\text{eff}$. We therefore introduce the scaled annihilation rate $\widetilde{Z}_\text{eff}$ that determines the energy dependence of $\overline Z_\text{eff}$ up to a multiplicative constant:
\begin{equation}
\widetilde Z_\text{eff}(\epsilon ) \equiv \frac{\overline Z_\text{eff}(\epsilon)}{\pi F\beta} = \kappa \sum_\nu \frac{\Delta(\epsilon-\omega_\nu+\varepsilon_b)}{k_\nu} , \label{eq:tilde_Zeff}
\end{equation}
where the sum is over the normal modes with $\omega_\nu>\varepsilon_b$.

To account for the effect of the presence of multiple conformers on $\widetilde Z_\text{eff}(\epsilon )$ in an $n$-alkane gas, we calculate $\widetilde Z_\text{eff}^{(i)}(\epsilon )$ for each conformer $i$ using Eq.~(\ref{eq:tilde_Zeff}) with the binding energy $\varepsilon_b^{(i)}$. The expected energy dependence of $\widetilde Z_\text{eff}$ is then given by the sum over the conformers
\begin{align}
\widetilde Z_\text{eff}(\epsilon)= \sum_i p_i \widetilde Z_\text{eff}^{(i)}(\epsilon) ,
\end{align}
where $p_i$ is given by Eq.~(\ref{eq:prob-dist}).
For those molecules for which we used random sampling to determine the average binding energy, we estimate $\widetilde Z_\text{eff} (\epsilon)$ simply by taking the arithmetic mean of the $\widetilde Z_\text{eff}^{(i)}(\epsilon)$ for the conformers in the sample.

Before presenting the results, we note that our aim here is not to provide a theoretical description of the measured annihilation rates, but to explore the effect that the presence of various conformers would have on the shapes of the measured annihilation rates. To this effect, we \textit{simulate} (rather than calculate from first principles or otherwise) the measured annihilation rates using $\widetilde Z_\text{eff}$. In principle, the analytical form of Eq.~(\ref{eq:zeff_bar_final}) can be used to calculate $\overline Z_\text{eff}$ and compare with experimental data, when values of the widths are available (see, e.g., Refs.~\onlinecite{Gribakin06} and \onlinecite{Gribakin17}), or to fit the experimental data by using the widths ratio $\beta _\nu$ as a fitting parameter, producing good agreement with the measurements [see, e.g., Ref.~\onlinecite{Gribakin09} (Fig. 3) and Ref.~\onlinecite{Danielson13}].

\begin{figure*}
\includegraphics[width=0.8\textwidth]{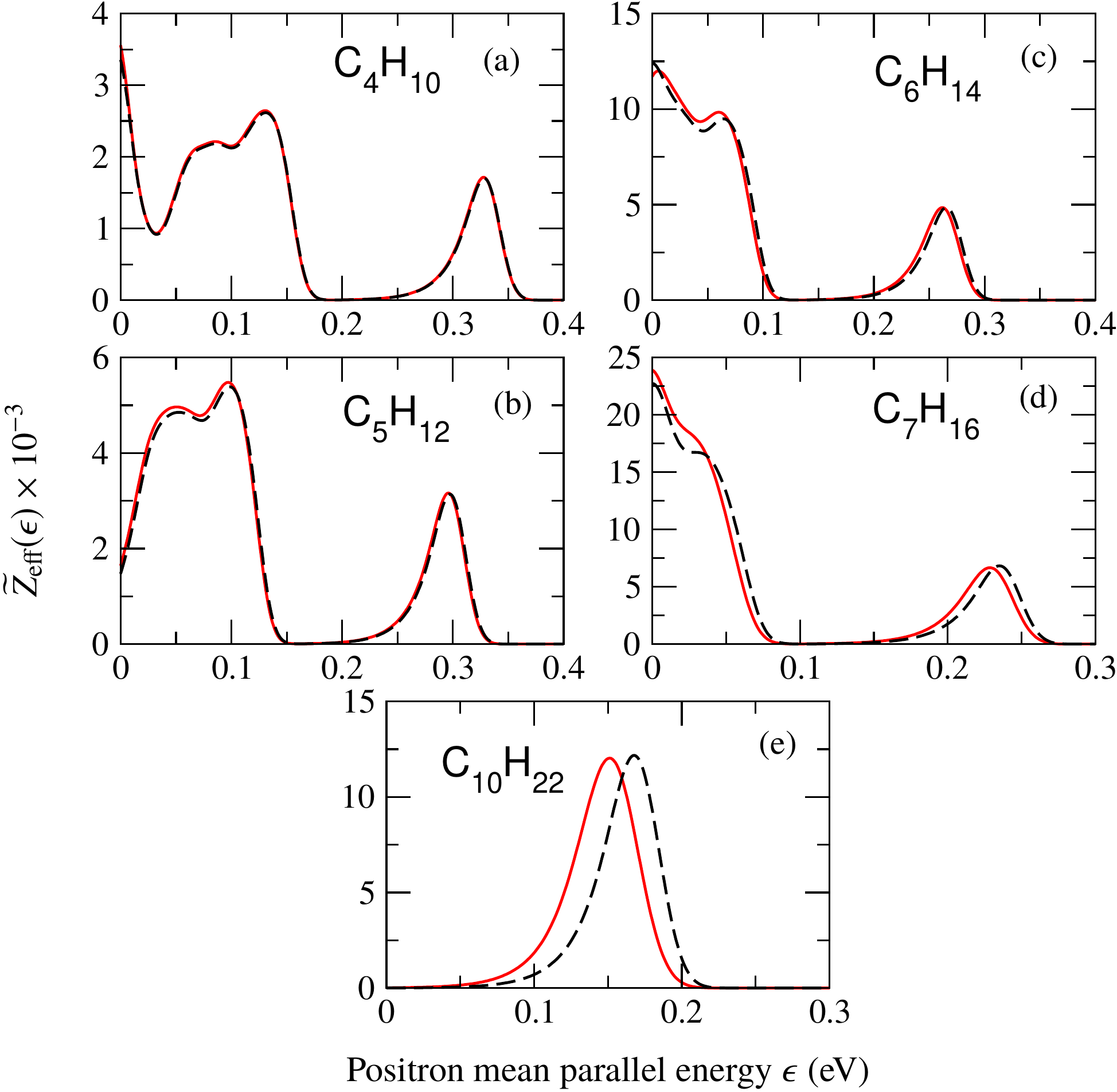}
\caption{\label{fig:Zeff_BGT}Predicted scaled annihilation rate $\widetilde Z_\text{eff}$ at $T=300$~K for (a) $n$-butane C$_4$H$_{10}$, (b) $n$-pentane C$_5$H$_{12}$, (c) $n$-hexane C$_6$H$_{14}$, (d) $n$-heptane C$_7$H$_{16}$, and (e) $n$-decane C$_{10}$H$_{22}$, for a BGT positron beam ($k_BT_\perp=20$~meV, $\sigma_z=10$~meV). Dashed black curves,  $\widetilde Z_\text{eff}^{(\text{all-}t)}$ for the all-$t$ conformer; solid red curves, the expected value $\widetilde Z_\text{eff}$ accounting for the presence of various conformers [using the full population with $\xi=1$ in (a)--(d), and using the random sample with $\Delta E_{tg}=40$~meV in (e)].}
\end{figure*}

Figure \ref{fig:Zeff_BGT} compares the energy dependence of $\widetilde Z_\text{eff}^{(\text{all-}t)}$ for the all-$t$ conformer, with that of the room-temperature ($T=300$~K) conformer-averaged $\widetilde Z_\text{eff}$, for $n=4$, 5, 6, 7, and 10, using the BGT-positron-beam parameters. 
%
%\blue{In Eq.~(\ref{eq:tilde_Zeff}), we only include those VFRs for which the molecule is excited into one of its $9n$ vibrational modes; we neglect the possibility of multimode vibrational excitations. It is known empirically that inclusion of the mode-based VFRs alone accurately describes the energy dependence of $Z_\text{eff}$, with the multimode VFRs playing a role in enhancing the overall magnitude of $Z_\text{eff}$.\footnote{This has been explained as a result of the positron first entering the bound state by transferring its excess energy into excitation of a vibrational mode. This so-called \textit{doorway} state of the positron-molecule complex is embedded in the dense spectrum of multimode vibrations. The doorway state decays (or ``spreads'') into multimode vibrational states.\cite{Gribakin04,Gribakin09}}}
The mode frequencies $\omega_\nu$ for each molecule were calculated using \textsc{q-chem}\cite{qchem,Gribakin04} for the all-$t$ conformer, and we use the same set for all conformers. The frequencies for various conformers are, strictly speaking, different. However, their overall distribution across the spectrum is quite similar, as discussed in Sec.~\ref{subsec:av_be}, e.g., the $2n+2$ C-H-stretch modes that give rise to a most prominent peak in $\Zeff$ are grouped around the 2800--2900~cm$^{-1}$ frequency range.

\begin{figure*}
\includegraphics[width=0.8\textwidth]{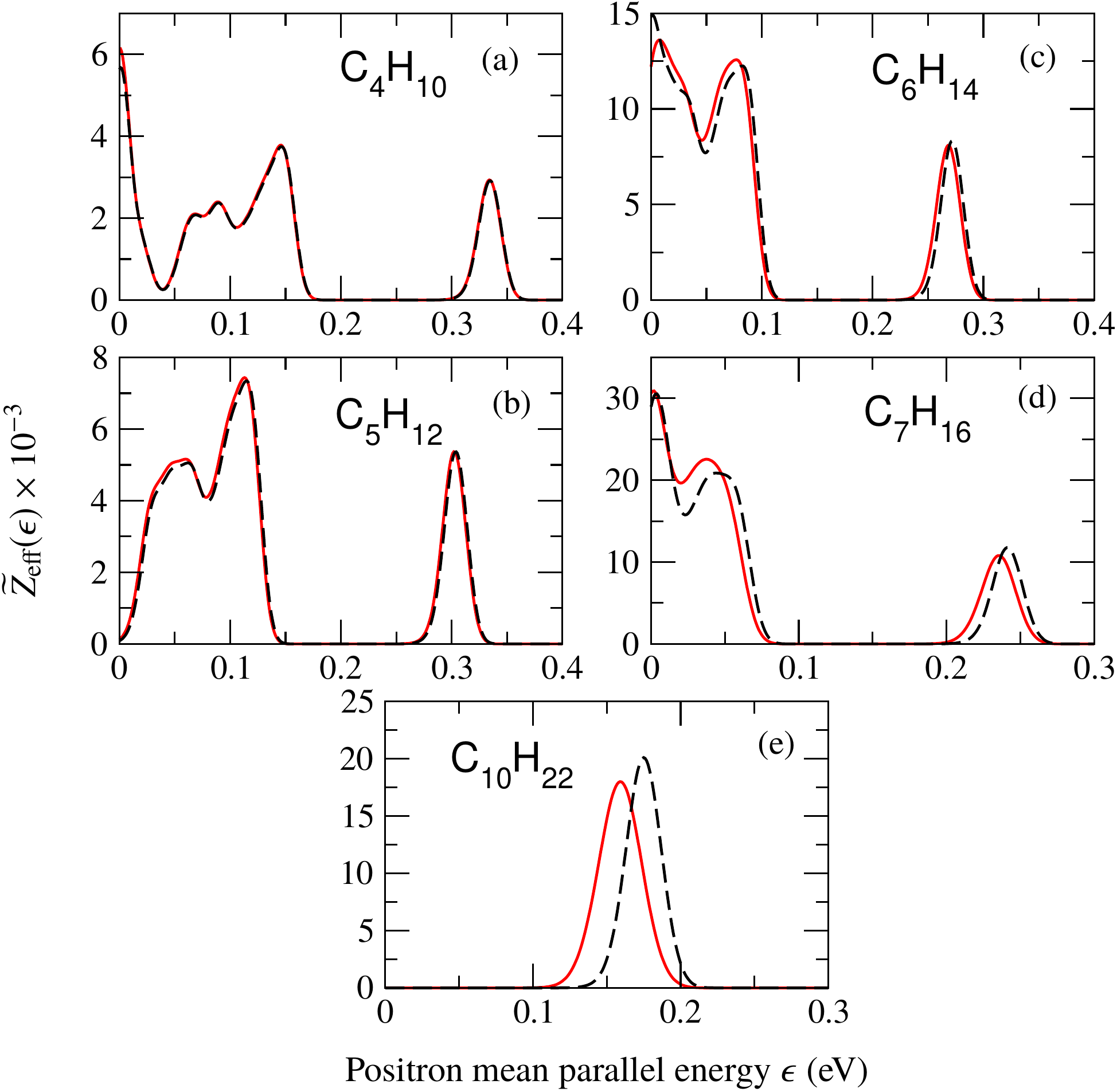}
\caption{\label{fig:Zeff_CBT}Predicted scaled annihilation rate $\widetilde Z_\text{eff}$ at $T=300$~K for (a) $n$-butane C$_4$H$_{10}$, (b) $n$-pentane C$_5$H$_{12}$, (c) $n$-hexane C$_6$H$_{14}$, (d) $n$-heptane C$_7$H$_{16}$, and (e) $n$-decane C$_{10}$H$_{22}$, for a CBT positron beam ($k_BT_\perp=5$~meV, $\sigma_z=8$~meV). Dashed black curves,  $\widetilde Z_\text{eff}^{(\text{all-}t)}$ for the all-$t$ conformer; solid red curves, the expected value $\widetilde Z_\text{eff}$ accounting for the presence of various conformers [using the full population with $\xi=1$ in (a)--(d), and using the random sample with $\Delta E_{tg}=40$~meV in (e)].}
\end{figure*}

For $n=4$--7, the calculations of $\widetilde Z_\text{eff}$ have been performed for all the conformers using the Hartree-Fock values of $E_i$ (i.e., with $\xi=1$, see Table~\ref{tab:n7conf}). For $n=10$, the calculation of $\widetilde Z_\text{eff}$ has been performed using the random sample of conformers with $\Delta E_{tg}=40$~meV (Table \ref{tab:conf_rand}). The corresponding scaled annihilation rates for a CBT-based positron beam are shown in Fig.~\ref{fig:Zeff_CBT}. The simplest feature to follow in these ``spectra'' is the high-energy peak that corresponds to the C-H-stretch vibrational modes. This peak is the tell-tale feature of the measured annihilation rates in all alkanes (and, in fact, most hydrocarbons) that allows for an accurate determination of the positron binding energies. The C-H stretch mode energies are $\omega_\nu\approx2900~\text{cm}^{-1}\approx0.36$~eV, and the corresponding annihilation rate peak is progressively downshifted following the increase of the positron binding energy with molecular size. 

For $n$-butane and $n$-pentane, the averaged $\widetilde Z_\text{eff}$ spectrum is essentially indistinguishable from the $\widetilde Z_\text{eff}^{(\text{all-}t)}$ spectrum for the all-$t$ conformer. However, for the larger molecules, some differences can be seen. For $n\geq6$, the C-H-stretch peak in the averaged $\widetilde Z_\text{eff}$ lies noticeably to the left of the peak in the $\widetilde Z_\text{eff}^{(\text{all-}t)}$ spectrum. This is a direct result of the increasing contribution of possible conformers with binding energies greater than that of the all-$t$ conformer for larger $n$.
The energy difference between the C-H-stretch peaks in $\widetilde Z_\text{eff}$ and $ \widetilde Z_\text{eff}^{(\text{all-}t)}$, for $n=4$, 5, 6, 7, and 10, for a BGT beam, is 0, 1, 4, 6, and 17~meV, respectively. The corresponding differences for a CBT beam are 1, 2, 4, 5, and 16~meV. For each $n$, these shifts are consistent with the corresponding difference between the average binding energy $\langle \varepsilon_b \rangle$ and the binding energy of the all-$t$ conformer, viz., 0.51, 1.42, 3.39, 6.00, and 15.4~meV.

In addition to the shift in position of the peak, a small amount of broadening of the resonance profile is observed. Table \ref{tab:broadening} shows the full width at half maximum (FWHM) of the C-H stretch peaks in $\widetilde Z_\text{eff}^{(\text{all-}t)}$ and the conformer average $\widetilde Z_\text{eff}$, for the BGT and CBT beams, for each $n$.
\begin{table}
\caption{\label{tab:broadening}FWHM (in meV) for the C-H stretch peaks in $\widetilde Z_\text{eff}^{(\text{all-}t)}$ for the all-$t$ conformer and conformer average $\widetilde Z_\text{eff}$, for the BGT and CBT positron beam parameters. Also shown are the values of the root-mean-squared deviation in the binding energy from the mean, $\Delta\varepsilon_b$ (in meV).}
\begin{ruledtabular}
\begin{tabular}{lccccr}
        & \multicolumn{2}{c}{BGT} & \multicolumn{2}{c}{CBT}  \\
        \cline{2-3} \cline{4-5}
$n$ & all-$t$ & average & all-$t$ & average & $\Delta\varepsilon_b$ \\
\hline
4 & 38 & 38 & 24 & 24 & 0.82 \\
5 & 38 & 38 & 24 & 25 & 1.7\phantom{0}  \\
6 & 38 & 40 & 24 & 25 & 3.4\phantom{0} \\
7 & 38 & 41 & 24 & 27 & 5.3\phantom{0} \\
10 & 41 & 48 & 28 & 34 & 8.3\phantom{0}
\end{tabular}
\end{ruledtabular}
\end{table}
For $n=4$--7, the FWHM for the all-$t$ conformer is 38~meV or 24~meV for the BGT or CBT, respectively (to the nearest meV). These widths are largely due to the positron-energy spread of the beam, but also due to the range covered by the C-H-stretch frequencies in $n$-alkanes (about 100~cm$^{-1}$ or 12 meV). Accounting for the thermal (room-temperature) population of all the conformers broadens the C-H-stretch peak. For $n=4$ (and $n=5$ for a BGT), the increase in the FWHM values is less than 1~meV.  For $n=5$ for a CBT, and for $n=6$ and $n=7$, the broadening is in the range 1--3~meV. For $n=10$, the broadening is 7~meV or 6~meV for a BGT or CBT, respectively.
It is noteworthy that for each $n$, the amount of broadening is almost the same for  a BGT and a CBT, although since the FHWM for the all-$t$ conformer is smaller for a CBT than for a BGT, the relative amount of broadening is greater for a CBT.
The amount of broadening can be compared with the root-mean-squared deviation of the binding energy from the mean, $\Delta\varepsilon_b$, which is given by
\begin{equation}\label{eq:rms}
(\Delta\varepsilon_b)^2=\sum_i p_i \big[ \varepsilon_b^{(i)} - \langle \varepsilon_b \rangle_T \big]^2=\sum_i p_i \big[ \varepsilon_b^{(i)} \big]^2 - \langle \varepsilon_b \rangle_T^2 ,
\end{equation}
with $T=300$~K. For $n=4$, 5, 6,  and 7, the sums in Eq.~(\ref{eq:rms}) are over all spectroscopically distinguishable conformers, with $p_i$ given by Eq.~(\ref{eq:prob-dist}). For $n=10$, the sums in Eq.~(\ref{eq:rms}) are over the 10 conformers in the random sample, with $p_i=\frac{1}{10}$. Table \ref{tab:broadening} shows the values of $\Delta\varepsilon_b$. We see that, typically, the amount of broadening is ${\sim}0.6\Delta\varepsilon_b$.

\begin{figure}
\includegraphics[width=\columnwidth]{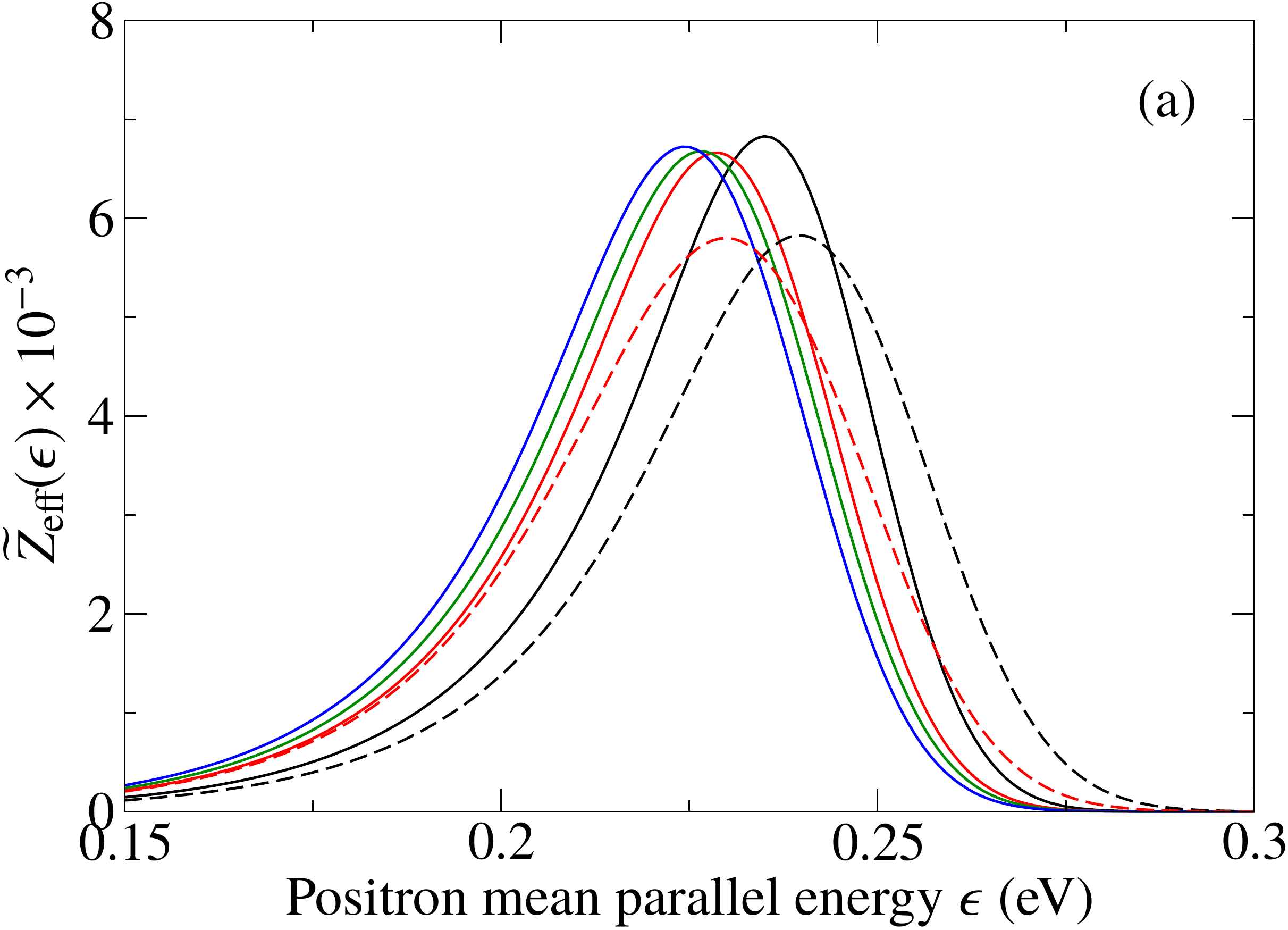} \\
\includegraphics[width=\columnwidth]{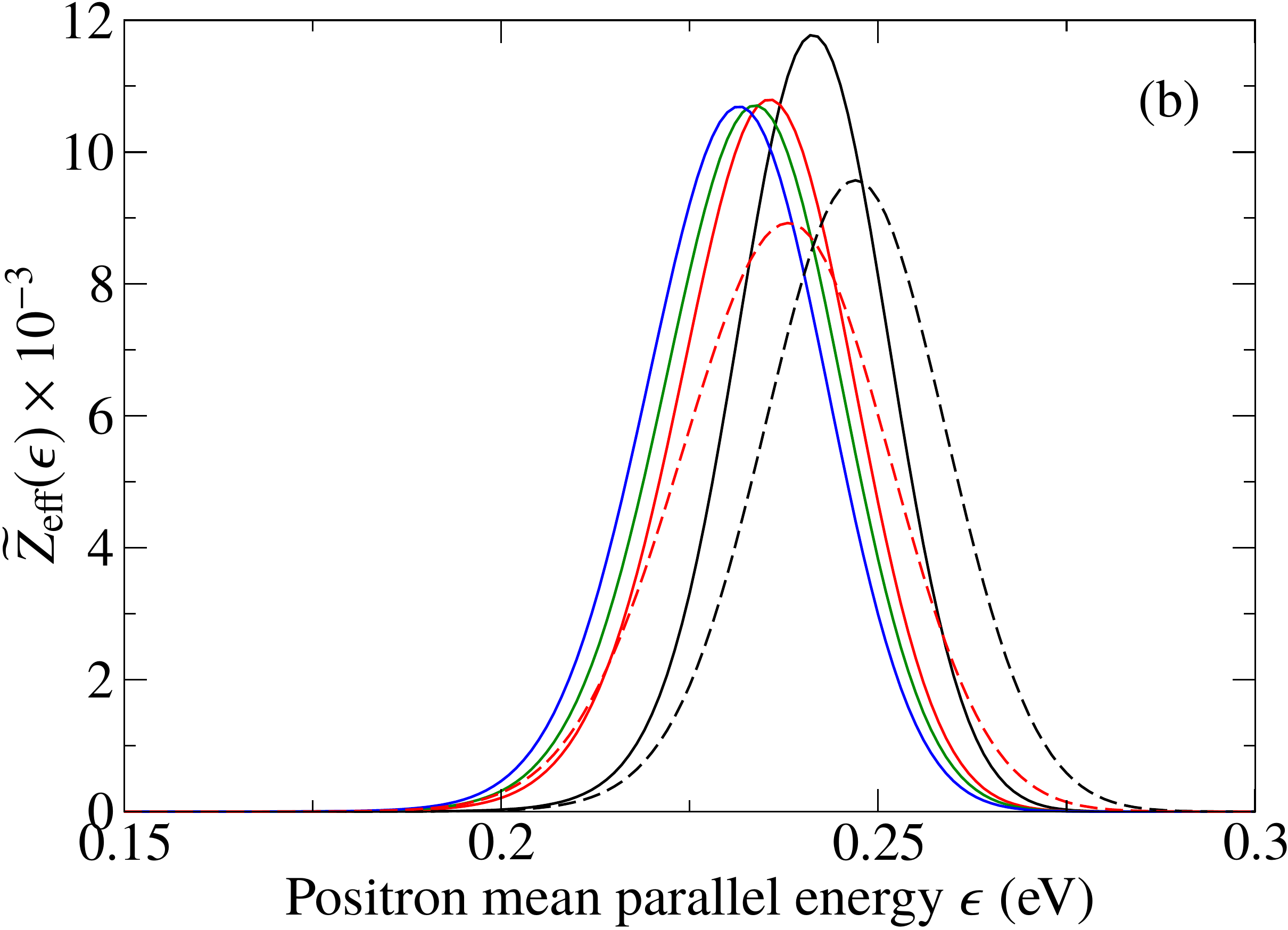}
\caption{\label{fig:Zeff_heptane_comp}Predicted $\widetilde{Z}_\text{eff}$ for the C-H-stretch peak in $n$-heptane C$_7$H$_{16}$ for (a) a BGT beam ($k_BT_\perp=20$~meV, $\sigma_z=10$~meV) and (b) a CBT beam ($k_BT_\perp=5$~meV, $\sigma_z=8$~meV).
Solid  curves are for calculations using the full population of conformers at $T=300$~K with Hartree-Fock-optimized geometry: black, $\widetilde{Z}_\text{eff}^{(\text{all-}t)}$ for the all-$t$ conformer; red, conformer-average $\widetilde{Z}_\text{eff}$ for $\xi=1$; green, $\langle\widetilde{Z}_\text{eff}\rangle$ for $\xi=0.75$; blue, $\langle\widetilde{Z}_\text{eff}\rangle$ for $\xi=0.5$. 
Dashed curves are for the random sample with \textsc{avogadro}-optimized geometry: black, $\widetilde{Z}_\text{eff}^{(\text{all-}t)}$; red, conformer average $\widetilde{Z}_\text{eff}$ for $\Delta E_{tg}=40$~meV.}
\end{figure}

An important question is whether the shifting and broadening of the C-H-stretch peak due to the presence of conformers is strongly affected if the values of the $E_i$ [for which we have so far used the Hartree-Fock values (i.e., $\xi=1$) for $n=4$--7, and $\Delta E_{tg}=40$~meV for $n=10$] are changed.
Figure \ref{fig:Zeff_heptane_comp} shows the C-H stretch peak for the all-$t$ conformer $\widetilde Z_\text{eff}^{(\text{all-}t)}$, and the conformer-average $\widetilde Z_\text{eff}$, for $n$-heptane. Values of $\widetilde Z_\text{eff}$ are shown for $\xi=1$ (as in Figs.~\ref{fig:Zeff_BGT} and \ref{fig:Zeff_CBT}), and also for $\xi=0.75$ and 0.5.
We observe that a reduction in the values of the single- and multi-\text{gauche} conformer energies $E_i$ does lead to a noticeable additional downshift in the position of the peak, due to increased populations of such conformers.  For $\xi=0.75$ and 0.5 for a BGT, the position of the peak is 8~meV and 11~meV lower, respectively, than for the all-$t$ conformer alone, as compared to being only 6~meV lower for $\xi=1$. The shifts for a CBT are similar. This is consistent with the increase in the average binding energy $\langle\varepsilon_b\rangle$ for reduced $E_i$: the values of $\langle\varepsilon_b\rangle$ for $\xi=0.75$ and 0.5 are 7.9~meV and 9.9~meV greater than the value of $\varepsilon_b$ for the all-$t$ conformer (see Fig.~\ref{fig:n=4-7_full_T_dep}).
Reduction of $E_i$ also causes additional broadening of the peak profile. However, this effect is barely discernible for $n$-heptane. For a BGT beam, the FWHM for both $\xi=0.75$ and 0.5 is the same as for $\xi=1$, to the nearest meV (viz., 41~meV);  for a CBT, the FWHM for $\xi=0.75$ is the same as for $\xi=1$, to the nearest meV (viz., 27~meV), and just 1~meV greater for $\xi=0.5$.

For completeness, Fig.~\ref{fig:Zeff_heptane_comp} also shows $\widetilde Z_\text{eff}^{(\text{all-}t)}$ and the conformer average $\widetilde Z_\text{eff}$, as calculated using the random sample at $T=300$~K for $n$-heptane (with $\Delta E_{tg}=40$~meV). The peaks of both curves lie to the right of the peaks of the corresponding curves calculated using the full population of conformers. This is a result of the different geometry optimization between the two methods. The binding energy for the all-$t$ conformer and the average binding energy calculated using the approximately (\textsc{avogadro}) optimized geometry are both a few meV smaller than the corresponding binding energy calculated using the fully (Hartree-Fock) optimized geometry (see Sec.~\ref{subsec:av_be}). 
For a BGT positron beam, the FWHM of the peak for the all-$t$ conformer is 44~meV, increasing to 47~meV for the conformer average $\widetilde Z_\text{eff}$ with $\Delta E_{tg}=40$~meV. For a CBT beam, the FWHM of $\widetilde Z_\text{eff}^{(\text{all-}t)}$ is 28~meV, increasing to 33~meV for $\widetilde Z_\text{eff}$ with $\Delta E_{tg}=40$~meV.
The random sample calculations thus predict about the same amount of broadening as the calculations using the full set of conformers; cf. Table \ref{tab:broadening}.

\section{Conclusions}

The ability of many (probably, most) polyatomic molecules to support a bound state for a positron is responsible for the orders-of-magnitude enhancement of the normalized annihilation rate $Z_\text{eff}$ above that of direct, ``in-flight'' annihilation.
For the alkane molecules, we have investigated the dependence of the positron binding energy on the molecule's constitution and geometry, i.e., $n$-alkane C$_n$H$_{2n+2}$ vs. cycloalkane C$_n$H$_{2n}$. For the $n$-alkanes, we have also determined the effect of the presence of conformers in a gas sample  on the binding energy and the resonant peaks in $Z_\text{eff}$. This has been done using a model-potential approach previously developed by the authors.\cite{Swann18,Swann19,Swann20}

For $n\leq6$, each cycloalkane has a smaller binding energy than the corresponding $n$-alkane; this was explained as being due to the former having two fewer H atoms and hence a smaller polarizability. The key role of the dipole polarizability is maintained for molecules whose size is smaller than the spatial extent of the weakly bound positron state. As the size of the molecule increases and the bound state become more compact, following the increase of its binding energy, the effect of molecular geometry takes over.
For $n=7$, the cycloalkane and $n$-alkane have almost the same binding energy, and for $n\geq8$, the cycloalkane has a larger binding energy than the $n$-alkane. This is because for cycloalkanes, the positron is on average closer to the regions of high electron density and to the many polarizable centres of the molecules.

For an $n$-alkane gas, accounting for the presence of conformers leads to average room-temperature values of the binding energy that are 2--12\% greater than that of the lowest-energy conformer (i.e., the extended all-$t$ conformer), for $n\leq16$. 
For $n=14$ and 16, we found that the average binding energies were in better agreement with experiment than the binding energies of the all-$t$ conformers.
The difference between the average binding energy and the binding energy of the lowest-energy conformer increases with temperature. For $n\geq6$, accounting for the presence of conformers also leads to a noticeable energy shift and broadening of the C-H--stretch peak in the energy-resolved $Z_\text{eff}$ measured with a trap-based positron beam.

The physics of trap-based positron beams and their use for measuring resonant annihilation rates are now well understood, and the use of cryogenic traps promises increased energy resolution. On the theory side, calculations of positron binding energies for polyatomic species have made much progress over the past few years. Taken together, these developments allow one to use positron annihilation as a probe of molecular structure and dynamics, e.g., the \textit{trans}-\textit{gauche} isomerization of alkanes. Combined with calculated binding energies, measurements at different gas temperatures can provide information on the \textit{trans}-\textit{gauche} energy differences that would be complementary to exisiting data. Conversely, where reliable conformer energies are available, positron resonant annihilation studies can be used as a thermometer for measuring internal molecular temperatures.

One source of uncertainty in our calculations is the values of the energies $E_i$ of various conformers relative the lowest-energy (all-\textit{trans}) conformer of the same molecule. We used Hartree-Fock values of $E_i$, but recognizing that the true values of $E_i$ may be smaller, we also showed the temperature dependence of the average positron binding energy using values of $E_i$ that were scaled from the Hartree-Fock values by a factor of 0.75 or 0.5.
As explained above, comparisons of measured positron binding energies for $n$-alkane gases with such calculations using adjustable values of $E_i$ could, in principle, provide a new probe of the values of $E_i$.

A possible extension of the work would be to investigate the dependence of the positron binding energy on the molecular conformation for cycloalkanes. However, we expect the variation in the positron binding energy between the various conformers of a cycloalkane to be smaller than for the corresponding $n$-alkane. Indeed, every conformer of a cycloalkane is a closed ring with approximately the same spatial extent, while for $n$-alkanes, having \textit{gauche} bonds in arbitrary positions leads to a variety of shapes and spatial extent.
For example, cyclohexane has two spectroscopically distinct conformers.\cite{Dragojlovic15} The binding energy for the lower-energy conformer (the \textit{chair}) is 75.62~meV, while the binding energy for the higher-energy conformer (the \textit{twist boat}) is 75.89~meV, a relative difference of only 0.4\%. 
This is in stark contrast to the binding energies for the conformers of $n$-hexane, which ranged from 87.23 to 101.0~meV (see Table \ref{tab:n7conf}).

In addition to calculating positron-molecule binding energies for the alkanes here and in Ref.~\onlinecite{Swann19}, we have so far also used the model-potential approach  to study positron binding to HCN,\cite{Swann18} and low-energy scattering and direct annihilation of positrons by several small diatomic molecules and methane.\cite{Swann20}
In future work, the method can extended to other classes of molecules, in particular those for which experimental data are available.\cite{Young07,Danielson12}  We will also calculate the annihilation $\gamma$-ray spectra for bound-state positrons. A wealth of experimental data on positron $\gamma$-ray spectra for a wide range of molecules has existed for a long time,\cite{Iwata97} and is only now beginning to be investigated theoretically.\cite{Ikabata18} However, in the context of the present work, it is unlikely that annihilation $\gamma$-ray spectra of different conformers or even isomers of alkanes will be much different from each other, given the primary importance of the bound-electron momentum distribution for the $\gamma$-ray Doppler shifts.\cite{Green12}

\begin{acknowledgments}
We are grateful to  J. R. Danielson, S. Ghosh, and C. M. Surko for useful discussions.
This work has been supported by the EPSRC UK, Grant No. EP/R006431/1.
\end{acknowledgments}

\section*{Data availability}
The data that support the findings of this study are available from the corresponding author upon reasonable request.

\appendix

\section{\label{sec:enumeration}Number of possible conformers of C$_{\boldsymbol{n}}$H$_{\boldsymbol{2n+2}}$, excluding those with $\boldsymbol{g^\pm g^\mp}$ pairs}

We wish to determine how many conformers exist for a given $n$-alkane C$_n$H$_{2n+2}$, excluding those which have adjacent $g^+$ and $g^-$ angles. Let $N= n-3$ be the number of dihedral angles in the molecule that can each be either $t$, $g^+$, or $g^-$. 
For $N=1$, there are exactly three conformers, viz., $t$, $g^+$, and $g^-$.
For $N\geq2$, the possible conformers can be listed by taking all of the conformers for the previous value of $N$ (i.e.,  $N-1$) and adding an extra angle to the end. If the previous terminal angle were $t$, then this new angle can be either $t$, $g^+$, or $g^-$, giving three new conformers. On the other hand, if the previous terminal angle were $g^+$ ($g^-$), then this new angle can be either $t$ or $g^+$ ($t$ or $g^-$), giving two new conformers. To illustrate this, Table \ref{tab:enumeration} lists all of the allowed conformers for $N=1$--3.
\begin{table}
\caption{\label{tab:enumeration}The allowed conformers for $N=1$--3. For  $N\geq2$, each row of conformers is obtained by taking a conformer for $N-1$ and adding an extra angle.}
\begin{ruledtabular}
\begin{tabular}{cccc}
$N$ & terminating with $t$ & terminating with $g^+$ & terminating with $g^-$ \\
\hline
1 & $t$ & $g^+$ & $g^-$ \\
2 & $tt$ & $tg^+$ & $tg^-$ \\
   & $g^+t$ & $g^+g^+$ \\
   & $g^-t$ & & $g^-g^-$ \\
3 & $ttt$ & $ttg^+$ & $ttg^-$ \\
& $tg^+t$ & $tg^+g^+$ \\
& $tg^-t$ & & $tg^-g^-$ \\
& $g^+tt$ & $g^+tg^+$ & $g^+tg^-$ \\
& $g^+g^+t$ & $g^+g^+g^+$ \\
& $g^-tt$ & $g^-tg^+$& $g^-tg^-$ \\
&  $g^-g^-t$ & & $g^-g^-g^-$   
\end{tabular}
\end{ruledtabular}
\end{table}

Let $f(N)$ denote the total number of allowed conformers, 
and let $f_0(N)$, $f_+(N)$, and $f_-(N)$ denote the number of these that terminate with a $t$, $g^+$, and $g^-$ angle, respectively. By definition, 
\begin{align}
f(N)=f_0(N)+f_+(N)+f_-(N). \label{eq:a1}
\end{align} 
When moving from $N-1$ to $N$, a $t$ angle can become the new terminal angle for all of the conformers for $N-1$. However, a $g^+$ ($g^-$) angle can only become the new terminal angle for those conformers that terminated with a $t$ or $g^+$ ($t$ or $g^-$). Hence, we have
\begin{align}
f_0(N) &= f(N-1) ,\label{eq:a2} \\
f_\pm(N) &= f_0(N-1) + f_\pm(N-1) , \label{eq:a3}
\end{align}
for $N\geq 2$. Also, by symmetry,
\begin{align}
f_+(N) = f_-(N) , \label{eq:a4}
\end{align}
for all $N$. Using Eqs.~(\ref{eq:a1}), (\ref{eq:a2}), and (\ref{eq:a4}), we find
\begin{align}
f(N+1) &= f_0(N+1) + 2f_+(N+1) \notag\\
&= f(N) + 2f_+(N+1). \label{eq:a5}
\end{align}
Then
\begin{align}
f(N+2) &= f(N+1) + 2f_+(N+2) \notag\\
&= f(N+1) + 2f_0(N+1) +2 f_+(N+1) \notag\\
&= f(N+1) + 2f(N) +2 f_+(N+1) , \label{eq:a6}
\end{align}
where Eqs.~(\ref{eq:a2}) and (\ref{eq:a3}) have been used.
By subtracting Eq.~(\ref{eq:a5}) from Eq.~(\ref{eq:a6}), we obtain the following recurrence relation:
\begin{align}
f(N+2) - 2 f(N+1)  - f(N) =0. \label{eq:a8}
\end{align}
Seeking solutions of Eq.~(\ref{eq:a8}) of the form $f(N)=C\alpha^N$, where $C$ and $\alpha$ are constants, yields the auxiliary equation
\begin{align}
\alpha^2 - 2\alpha - 1 = 0,
\end{align}
whose solutions are $\alpha=1\pm\sqrt{2}$. Thus, the general solution of Eq.~(\ref{eq:a8}) is
\begin{align}
f(N) = C_1 \big(1-\sqrt{2}\big)^N + C_2 \big(1+\sqrt{2}\big)^N .
\end{align}
Using the conditions $f(1)=3$, $f(2)=7$ (see Table~\ref{tab:enumeration}) gives $C_1=(1-\sqrt{2})/2$, $C_2=(1+\sqrt{2})/2$. Hence,
\begin{align}
f(N) = \frac12 \left[ \big(1-\sqrt{2}\big)^{N+1} +  \big(1+\sqrt{2}\big)^{N+1} \right] ,
\end{align}
which, upon replacing $N$ by $n-3$, is the LHS of Eq.~(\ref{eq:restricted_total}). The RHS of Eq.~(\ref{eq:restricted_total}) is obtained by expanding the quantities in parentheses.

%\bibliography{conformers_bib}


\end{document}